\documentclass[epj]{svjour}

\usepackage{graphics}
\usepackage{latexsym}
\usepackage{graphics}
\usepackage{graphicx} 
\usepackage{dcolumn}
\usepackage{epsfig}
\usepackage{amssymb} 
\usepackage{amsmath} 
\usepackage{rotating} 
\usepackage{color} 
\newcommand{\gammap}{\dot{\gamma}}

\newcommand{\seuil}{\sigma_{\scriptscriptstyle 0}}

\newcommand{\gammapeff}{\gammap_{\hbox{\rm\scriptsize eff}}}
\begin{document}

\title{Towards local rheology of emulsions under Couette flow using Dynamic Light Scattering}

\author{Jean-Baptiste Salmon \and Lydiane B\'ecu \and S\'ebastien Manneville \and Annie Colin}

\institute{Centre de Recherche Paul Pascal, Avenue Schweitzer,
33600 Pessac, FRANCE}

\date{\today}

\abstract{
We present local velocity measurements in emulsions under shear using heterodyne
Dynamic Light Scattering. Two emulsions are studied: a dilute system of volume
fraction $\phi=20$~\% and a concentrated system with $\phi=75$~\%. Velocity profiles
in both systems clearly show the presence of wall slip. We investigate the evolution
of slip velocities as a function of shear
stress and discuss the validity of the corrections for wall slip
classically used in rheology. Focussing on the bulk flow, we show that
the dilute system is Newtonian and that the concentrated
emulsion is shear-thinning. In the latter case, the curvature of the velocity profiles
is compatible with a shear-thinning exponent of 0.4 consistent with global rheological
data. However, even if individual
profiles can be accounted for by a power-law fluid (with or without a
yield stress), we could not find a fixed set of parameters 
that would fit the whole range of applied shear rates.
Our data thus raise the question of the definition of a global flow
curve for such a concentrated system. These results show
that local measurements are a crucial
complement to standard rheological tools. They are discussed in light of
recent works on soft glassy materials.
}
 
\authorrunning{J.-B. Salmon \textit{et al.}}
\titlerunning{Towards local rheology of emulsions under Couette flow using DLS}
\maketitle

\section{Introduction}

Emulsions are mixtures of two immiscible fluids. They are usually formed
by mechanical mixing and consist of  droplets of one fluid dispersed in
the other. They are metastable so that the mean size of the droplets
tends to increase with time. Emulsions belong to a wide class of
non-equilibrium systems such as foams
that rearrange and coarsen with time \cite{Larson:1999}. The use of surfactants or polymers
increases the characteristic time for coarsening of emulsions from a few
seconds to a few years and make them very attractive for applications
such as road building, pharmacology or food processing. They constitute
very interesting tools to entrap hazardous solvents, to carry additives
or to change the rheological properties of a sample. Indeed, their
rheological characteristics may be tuned by increasing  the liquid fraction
of the dispersed phase.

When the liquid fraction is low, the droplets
of the emulsion are Brownian and the emulsion is a Newtonian fluid
whose viscosity is close to that of the continuous phase. When
the liquid fraction is increased, less space is available to each
particle. The droplets are pressed against each other, their shape
are distorted, and energy is stored at their interfaces. For small
deformations, the droplets resist shear elastically. Such elasticity
results from the energy stored by additional deformation of their shape
induced by the applied strain. For large deformations, they flow
irreversibly.

Mason {\it et al.} \cite{Mason:1996a} have shown that this change of regime occurs above a
critical stress and that, for concentrated emulsions in the neighbourhood of this critical stress, the flow may
become inhomogeneous. In Ref.~\cite{Mason:1996a}, the homogeneity of the
flow was determined by painting a stripe on the exposed surface of the
emulsion before shearing and by following its evolution under shear.
Moreover, the analysis of the rheological data may be complicated by the
existence of slippage of the emulsion at the walls \cite{Barnes:1994}.

Due to all
these difficulties, local measurements of the velocity are needed to
characterize the flow of the emulsion once it has yielded. Recently,
Coussot {\it et  al.} \cite{Coussot:2002} measured velocity profiles of various
colloidal systems such as gels, laponite solutions or industrial
emulsions by imaging the flow field with Magnetic Resonance Imaging (MRI).
They described the yield stress
phenomenon as an abrupt transition between two distinct states:  a
liquid-like state that flows, and a solid-like state that remains jammed and does
not flow. The rheological behaviour of the liquid
state was modelled by that of a power-law fluid.

In this article, we present measurements of velocity profiles in sheared emulsions
using Dynamic Light Scattering (DLS) in the heterodyne geometry.
Our setup allows us to perform both global rheological measurements
and local velocity measurements simultaneously.
We study two different oil-in-water
emulsions: a dilute emulsion of volume fraction $\phi = 20~\%$ and a concentrated one at $\phi = 75~\%$.
In both cases, we get a quantitative measure of wall slip
and show the importance of this
phenomenon. We point out that corrections for slippage like those proposed by
Mooney and classically used in
rheological experiments are indeed well suited for concentrated emulsions
but may fail for dilute systems. In the case of the
concentrated emulsion, we show the existence of two regimes 
separated by an abrupt transition between a jammed state and a flowing
phase like Coussot {\it et al} \cite{Coussot:2002}. However, in our case, no clear
analytical flow curve could be obtained for
the whole range of shear rates investigated in the present study.

In the next Section of this article, we describe the method used for preparing the samples. The third Section briefly
recalls our experimental technique and setup for measuring the local velocity of a complex fluid in a Couette flow
\cite{Salmon:2002pp}.
We then present the velocity profiles obtained in the
two emulsions under shear. Finally, we discuss both the slippage and the yield stress
phenomena in our emulsions.

\section{Preparation of the emulsions under study}
\label{s.preparation}

We prepare a crude emulsion of fixed composition
by gently shear-mixing 400~g of silicone oil (Polydymethyl Siloxane of viscosity
135~Pa.s from Rhodia), 30.3~g of glycerol, 30.3~g of water and
9.9~g of surfactant (Tetradecyl Trimethyl Ammonium Bromide from Aldrich).
The aqueous phase is thus composed of a 14~\% wt. solution of TTAB in a 1:1 mixture
of water and glycerol.
The resulting polydisperse viscoelastic premixed emulsion 
of large droplets is sheared within a narrow gap in a laminar regime \cite{Mason:1996b}.

\begin{figure}[htbp]
\begin{center}
\scalebox{2}{\includegraphics{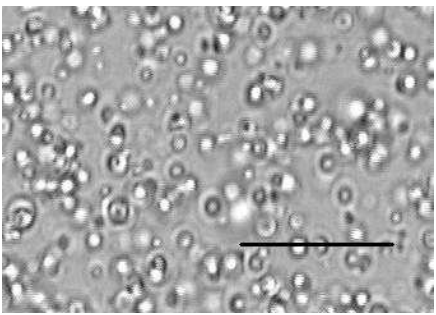}}
\end{center}
\caption{Photograph of the emulsion under study. The concentrated emulsion was diluted in order
to visualize individual droplets using a microscope
(magnification $\times$50). The black bar represents 20~$\mu$m.}
\label{photoemul}
\end{figure}
The mother emulsion obtained after shearing is composed of monodisperse
silicone oil droplets of diameter $2~\mu$m in the water--glycerol--surfactant mixture
(see Fig.~\ref{photoemul}).
 Then, by dilution, we set the surfactant
concentration within the water--glycerol phase to $1~\%$
 in mass and concentrate the droplets by centrifugation to a volume fraction of $\phi=75~\%$, 
such that droplets are compressed one against another leading to flat films at each contact.
 We chose this surfactant concentration in order to ensure a good stability to the emulsion. 
This concentration is high enough to prevent coalescence and small enough to avoid flocculation by depletion. 
We prepare another emulsion with a volume fraction $\phi= 20~\%$ by diluting the concentrated one 
in the water--glycerol--surfactant solution. The surfactant concentration is kept equal to $1~\%$.

The use of a water--glycerol mixture allows us to match the optical index $n$ of the continuous phase 
of the emulsion to that of the silicone oil ($n=1.40$). 
The obtained emulsions are thus nearly transparent and do not present any significant multiple light scattering. 
The emulsion is then carefully loaded into the transparent Couette cell of a rheometer (see below). 
For the whole range of applied shear rates,
we check by optical microscopy that the structure of the emulsion and the
size of the droplets are not affected by shear and that no noticeable coalescence
takes place.

\section{Experimental setup for heterodyne DLS in Couette flows}
\label{s.setup}

In order to measure the local velocity of emulsions in Couette flows, we used the experimental
technique based on heterodyne DLS described at length in
a related work \cite{Salmon:2002pp}. In this Section, we briefly recall the basics of heterodyne DLS and the
setup developed in our laboratory to perform local velocity measurements. The reader is referred to
Ref.~\cite{Salmon:2002pp} for more details on the optics of this setup and for a complete discussion
on the resolution of heterodyne DLS.
\begin{figure}[htbp]
\begin{center}
\scalebox{1}{\includegraphics{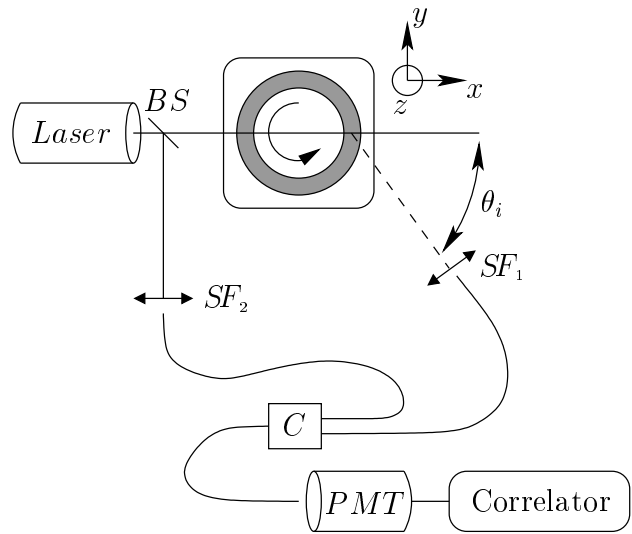}}
\caption{Simplified heterodyne DLS setup. $BS$ denotes a beam splitter, $SF$ spatial filters, and $C$ the device
coupling optical fibers and used to perform the interference between the scattered light and the reference beam.}
\label{setup}
\end{center}
\end{figure}

Local velocity measurements using heterodyne DLS relie on the detection of the Doppler frequency shift
associated with the motion of the scatterers inside a small scattering volume $\mathcal{V}$
\cite{Berne:1995,Ackerson:1981,Gollub:1974}.
In classical heterodyne setups, light scattered by the sample under study is collected along 
a direction $\theta$. The scattering volume is defined as the intersection between the incident beam and 
the scattered beam. The scattered electric field is then made to interfere with a reference beam. 
Finally, light resulting from the interference is sent to a photomultiplier tube (PMT)
and the auto-correlation function of the intensity is computed
using an electronic correlator. The originality of our setup lies in the use of single mode fibers to
collect scattered light, to perform the interference, and to carry light to the PMT.
This allows for great flexibility when choosing the scattering angle.
Figure~\ref{setup} sketches the main features of such a setup.

\begin{figure}[htbp]
\begin{center}
\scalebox{0.6}{\includegraphics{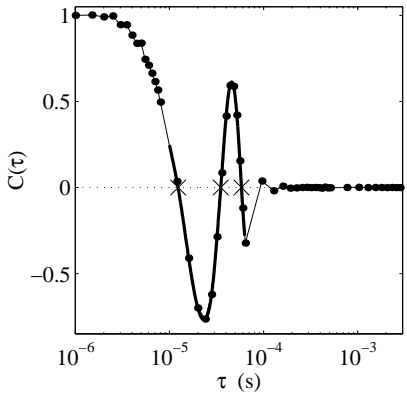}}
\end{center}
\caption{Experimental heterodyne correlation function ($\bullet$) recorded on the
concentrated emulsion at $\gammap=10$~s$^{-1}$.
The thick line shows the portion of the heterodyne
function from which the frequency shift $\mathbf{q}\cdot\mathbf{v}$
is calculated.}
\label{exphete}
\end{figure}

When the scattering volume $\mathcal{V}$ is submitted to a shear flow,
it can be shown that the correlation function $C(\tau)$ is an oscillating
function of the time lag $\tau$ modulated by a slowly decreasing envelope.
The frequency of the oscillations in $C(\tau)$
is exactly the Doppler shift $\mathbf{q}\cdot\mathbf{v}$, where $\mathbf{q}$
is the scattering wavevector and $\mathbf{v}$ is the local velocity averaged
over the size of the scattering volume which is about 100~$\mu$m in our experiments.
Figure~\ref{exphete} shows a typical correlation function measured on a sheared emulsion.
The frequency shift $\mathbf{q}\cdot\mathbf{v}$ is recovered by interpolating a
portion of $C(\tau)$ and looking for the zero crossings. Error bars on such measurements
are obtained by varying the number of zeros taken into account in the analysis. Typical
uncertainties are about 5~\%.

Couette flows are generated between two cylinders in a transparent Mooney-Couette cell.
The radius of the rotating inner cylinder (``rotor'') is $R_{\scriptscriptstyle 1}=22$~mm and the radius of the fixed 
outer cylinder (``stator'') is $R_{\scriptscriptstyle 2}=25$~mm leaving a gap $e=R_{\scriptscriptstyle 2}-R_{\scriptscriptstyle 1}=3$~mm between the two cylinders. The
length of the cylinders is $H=30$~mm. Both cylinders are in Plexiglas and have smooth surfaces
so that wall slip is likely to occur in our experiments. The
rotation of the moving cylinder is imposed by a classical rheometer (TA Instruments AR1000) operated
in the ``imposed shear rate'' mode.
This rheometer measures both the torque $\Gamma$ on the inner cylinder
and its angular velocity $\Omega$ in
real-time. From $\Omega$ and $\Gamma$, it also 
computes a {\it global} shear rate $<\gammap>$ and shear stress $<\sigma>$ using
the formulas:
\begin{eqnarray}
<\gammap> &=& \frac{R_{\scriptscriptstyle 1}^2+R_{\scriptscriptstyle 2}^2}{R_{\scriptscriptstyle 2}^2-R_{\scriptscriptstyle 1}^2}\,\Omega\, ,\label{e.gammarheo}
\\
<\sigma> &=& \frac{R_{\scriptscriptstyle 1}^2+R_{\scriptscriptstyle 2}^2}{4\pi H R_{\scriptscriptstyle 1}^2 R_{\scriptscriptstyle 2}^2}\,\Gamma\, .
\label{e.sigmarheo}
\end{eqnarray}
In the following, the results will be presented in terms of $<\gammap>$ and
$<\sigma>$ which will simply be noted $\gammap$ and $\sigma$,
as is usually done in rheology. Using Eq.~(\ref{e.gammarheo}), the
velocity of the moving wall $v_{\scriptscriptstyle 0}=R_{\scriptscriptstyle 1}\Omega$ is given by:
\begin{equation}
v_{\scriptscriptstyle 0}=\frac{R_{\scriptscriptstyle 1}(R_{\scriptscriptstyle 1}+R_{\scriptscriptstyle 2})}{R_{\scriptscriptstyle 1}^2+R_{\scriptscriptstyle 2}^2}\,\gammap e\, .
\label{e.gammacorr}
\end{equation}
Thus, in the Couette geometry used in the present work, one has to keep in mind that
$v_{\scriptscriptstyle 0}\approx0.93\,\gammap e$
so that $v_{\scriptscriptstyle 0}$ differs from $\gammap e$ by about 7~\%.

The temperature of the sample is controlled within $\pm\,0.1^{\circ}$C using
a water circulation around the cell.
The rheometer sits on a mechanical table whose displacements are controlled by a computer.
Three mechanical actuators allow us to move the rheometer in the $x$, $y$, and
$z$ directions with a precision of $1~\mu$m. Once $y$ and $z$ are set so that the incident
beam is normal to the cell surface, velocity profiles are measured by moving the mechanical
table in the $x$ direction by steps of 50 or 100~$\mu$m. 

As discussed in Ref.~\cite{Salmon:2002pp}, going from $\mathbf{q}\cdot\mathbf{v}$ as a function
of the table position $x_t$ to the velocity profile $v(x)$ (where $x$ is the radial position of the
scattering volume $\mathcal{V}$ within the sample) requires a careful calibration procedure. Indeed,
refraction effects due to the curved geometry of the Couette cell lead to significant differences
between the angle $\theta_i$ imposed by the operator
and the actual scattering angle $\theta$ in the sample, and between
the table displacement $\delta x$ and the actual displacement $\delta x'$ of $\mathcal{V}$.
By using a Newtonian suspension of latex spheres
in a water--glycerol mixture (optical index $n=1.40$) at various known shear rates,
we showed that for $\theta_i=35^\circ$, $\theta=f_\theta\,\theta_i$ and $\delta x'=f_x\,\delta x$ with
$f_\theta\approx0.79$ and $f_x\approx1.13$.

When loading the cell with the emulsions, great care is taken to ensure that the position of
the Couette cell does not change between the calibration with the latex suspension
and the actual experiments on emulsions. Since the optical index of the fluid is $n=1.40$
in both cases, the above values of $f_x$ and $f_\theta$ found with the latex suspension
are used to convert raw data $(\mathbf{q}\cdot\mathbf{v})(x_t)$ into
velocity profiles $v(x)$ in the case of our emulsions. Finally,
we chose to accumulate the correlation functions over 1~min so that a full velocity profile
with a resolution of 100~$\mu$m takes about 30~min to complete. Note that our setup allows us
to record rheological data and the local velocity at the same time on the same sample.
In all the cases described below, we checked that steady state was reached from the curves
$\gammap(t)$ and $\sigma(t)$.

\section{Velocity profiles in sheared emulsions}

\subsection{Dilute emulsion ($\phi=20~\%$)}
\begin{figure}[htbp]
\begin{center}
\scalebox{0.45}{\includegraphics{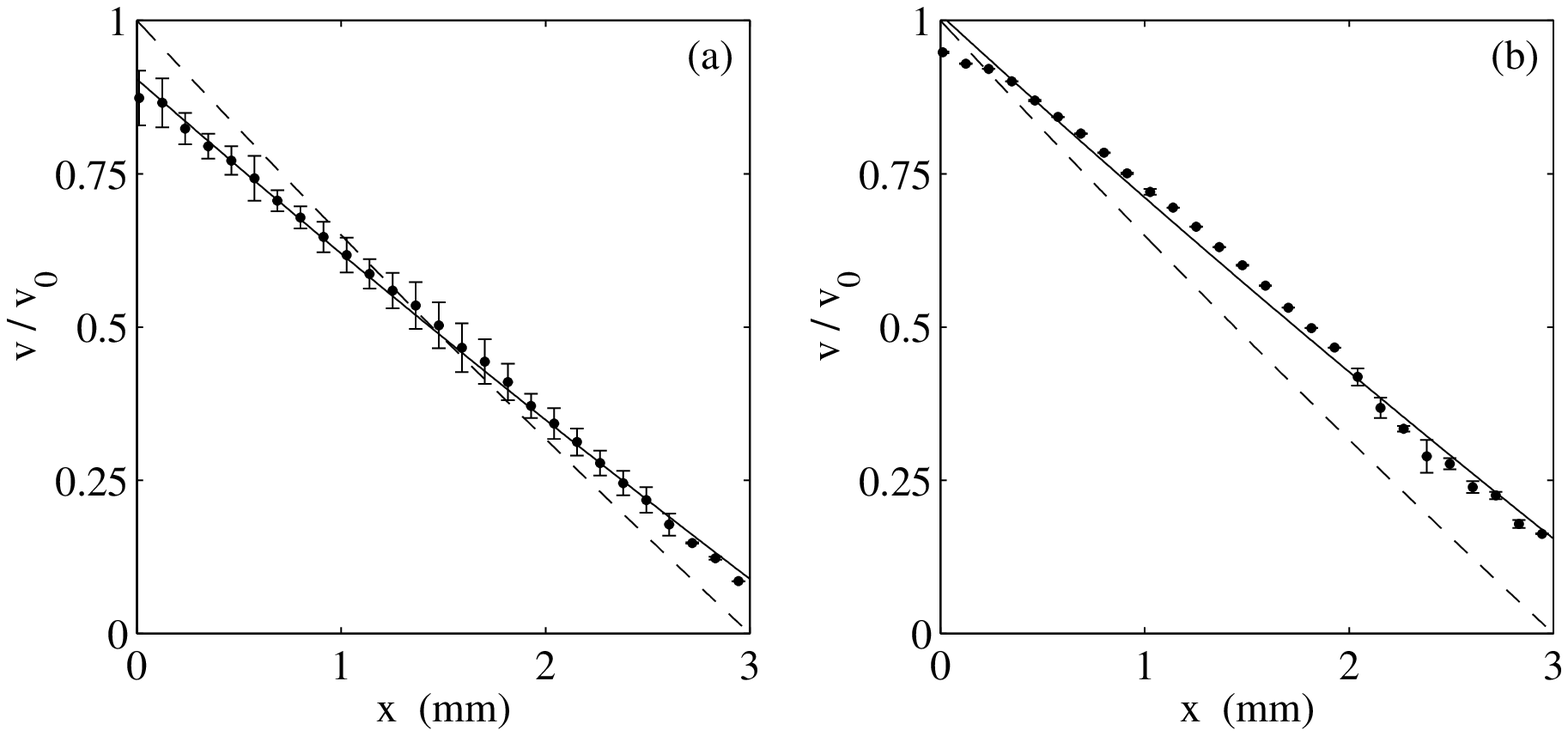}}
\end{center}
\caption{\label{emu20_01_02}Velocity profiles in
the 20~\% emulsion for (a) $\gammap=0.1$~s$^{-1}$
and (b) $\gammap=0.2$~s$^{-1}$.
The solid curves are the best fits to the data using Eq.~(\ref{vfit}) with $n=1$.
The dashed lines
represent the velocity profiles expected for a Newtonian fluid.
}
\end{figure}

\begin{figure}[htbp]
\begin{center}
\scalebox{0.45}{\includegraphics{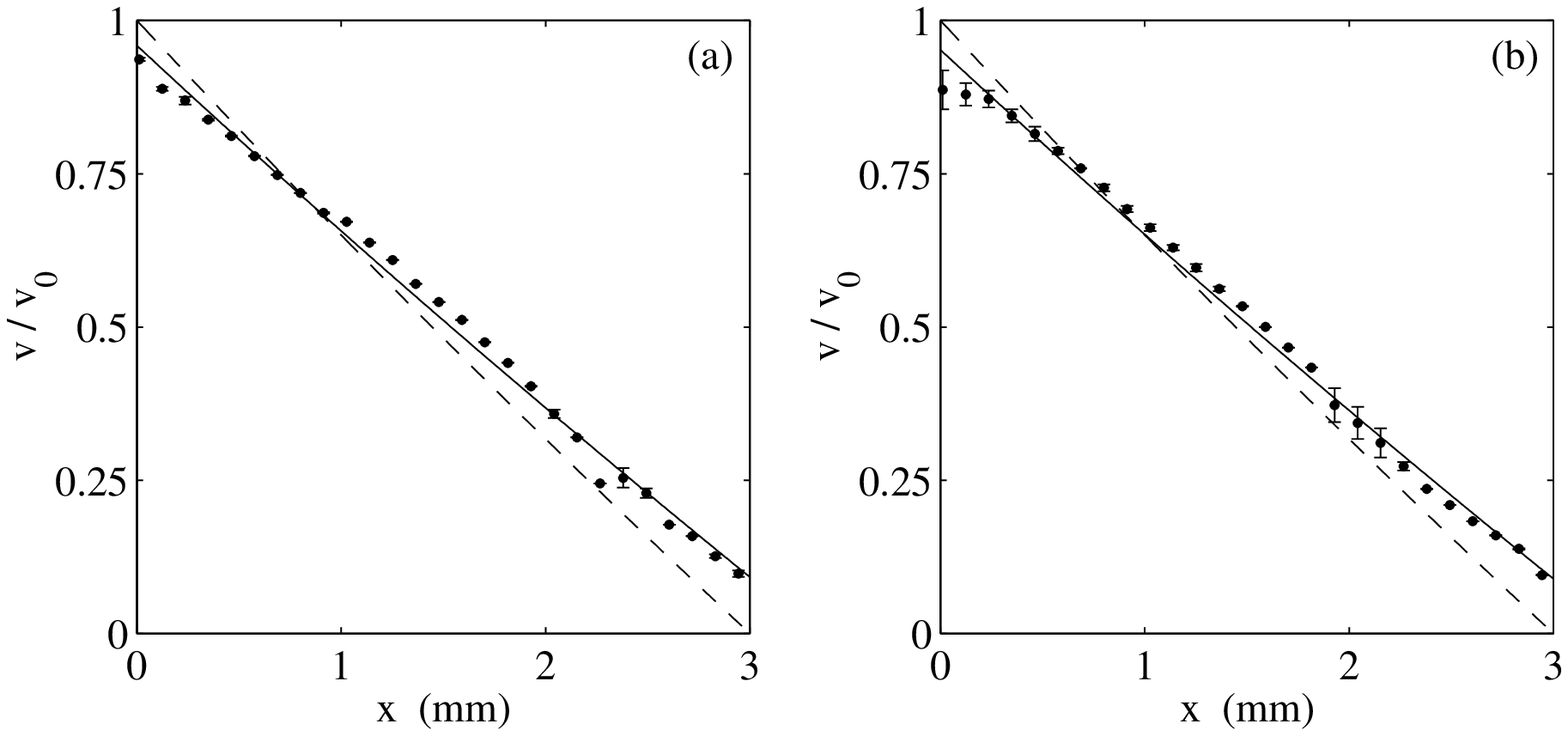}}
\end{center}
\caption{\label{emu20_04_1}Velocity profiles in
the 20~\% emulsion for (a) $\gammap=0.4$~s$^{-1}$
and (b) $\gammap=1$~s$^{-1}$.
The solid curves are the best fits to the data using Eq.~(\ref{vfit}) with $n=1$.
The dashed lines
represent the velocity profiles expected for a Newtonian fluid.
}
\end{figure}

\begin{figure}[htbp]
\begin{center}
\scalebox{0.45}{\includegraphics{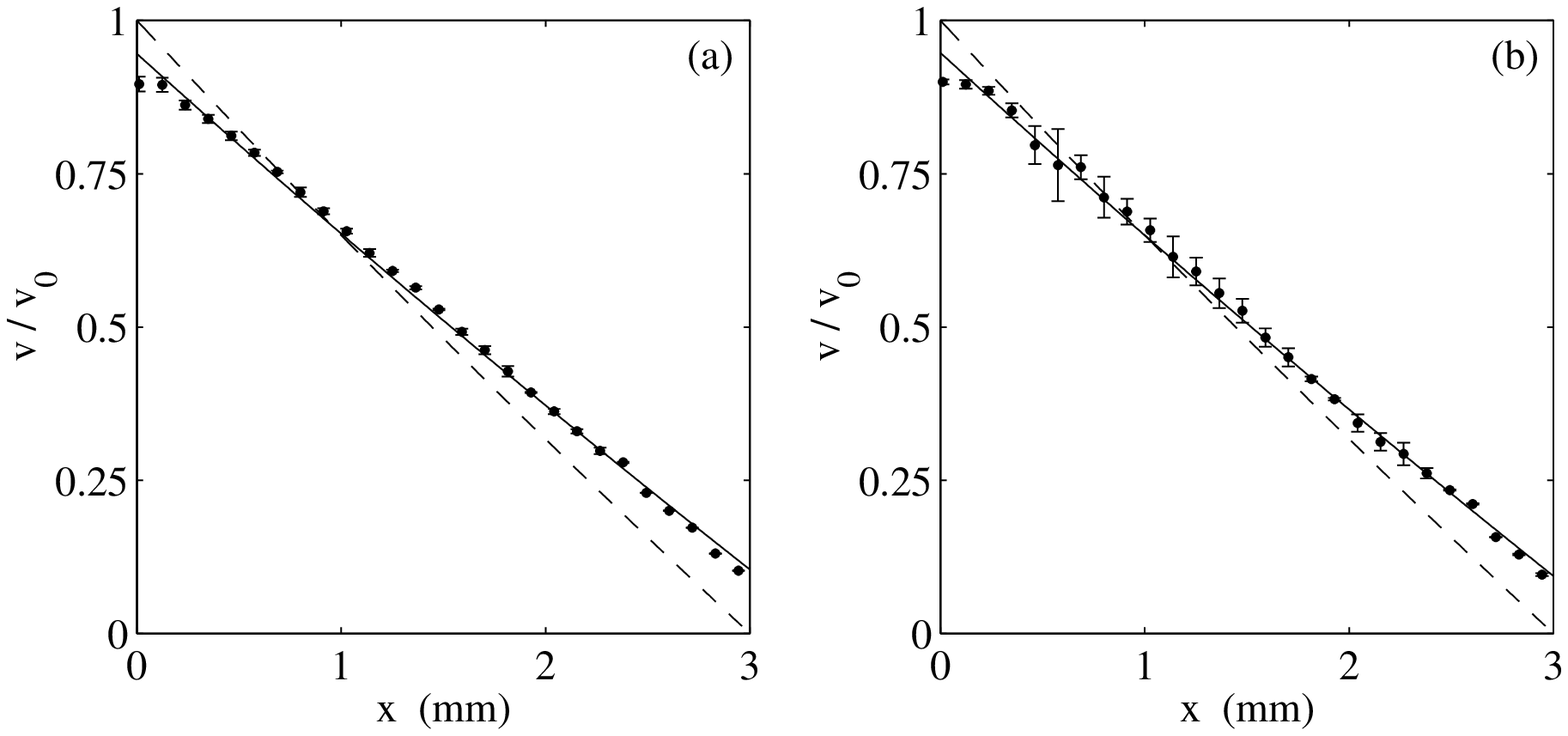}}
\end{center}
\caption{\label{emu20_2_5}Velocity profiles in
the 20~\% emulsion for (a) $\gammap=2$~s$^{-1}$
and (b) $\gammap=5$~s$^{-1}$.
The solid curves are the best fits to the data using Eq.~(\ref{vfit}) with $n=1$.
The dashed lines
represent the velocity profiles expected for a Newtonian fluid.
}
\end{figure}

\begin{figure}[htbp]
\begin{center}
\scalebox{0.45}{\includegraphics{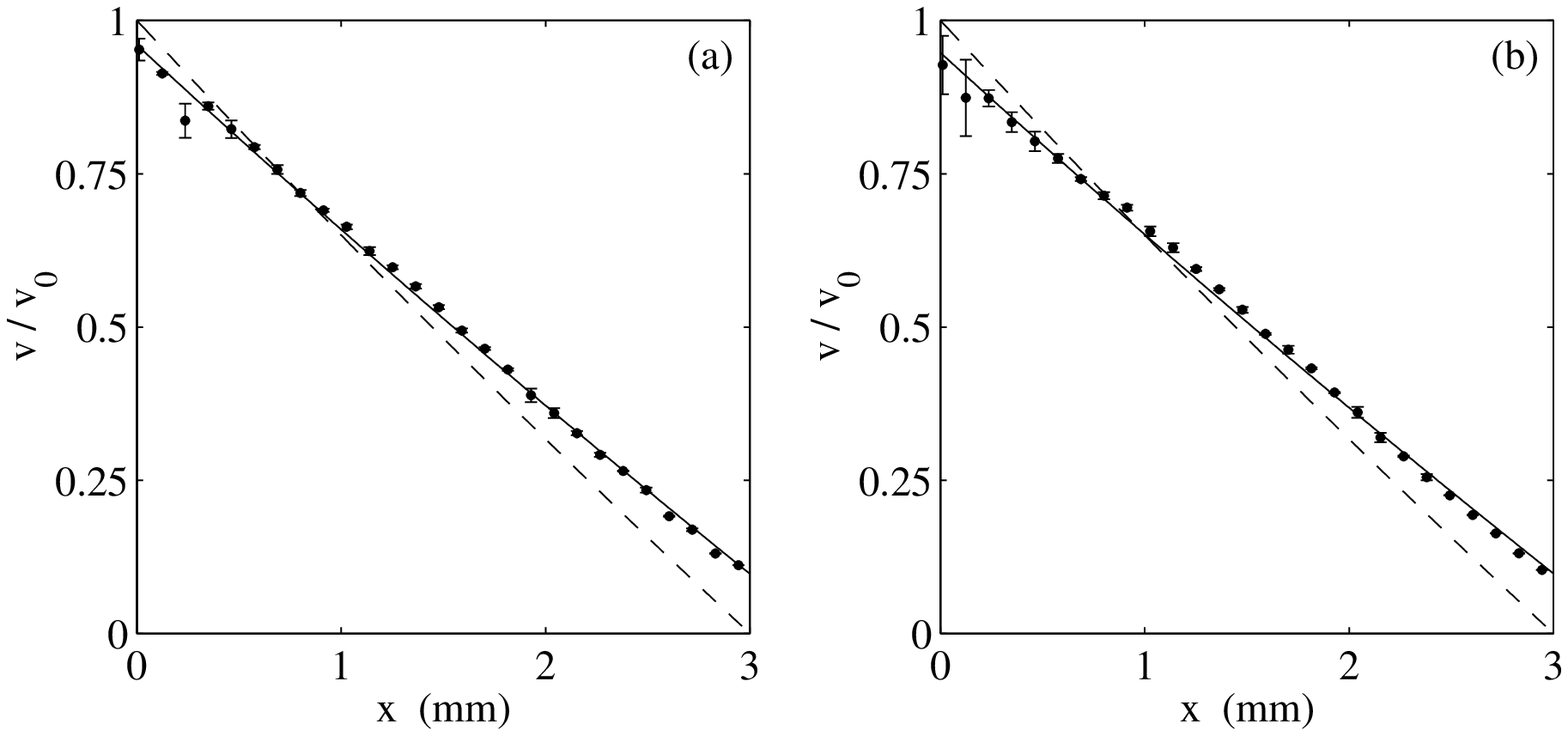}}
\end{center}
\caption{\label{emu20_10_15}Velocity profiles in
the 20~\% emulsion for (a) $\gammap=10$~s$^{-1}$
and (b) $\gammap=15$~s$^{-1}$.
The solid curves are the best fits to the data using Eq.~(\ref{vfit}) with $n=1$.
The dashed lines
represent the velocity profiles expected for a Newtonian fluid.
}
\end{figure}
\begin{figure}[htbp]
\begin{center}
\scalebox{0.6}{\includegraphics{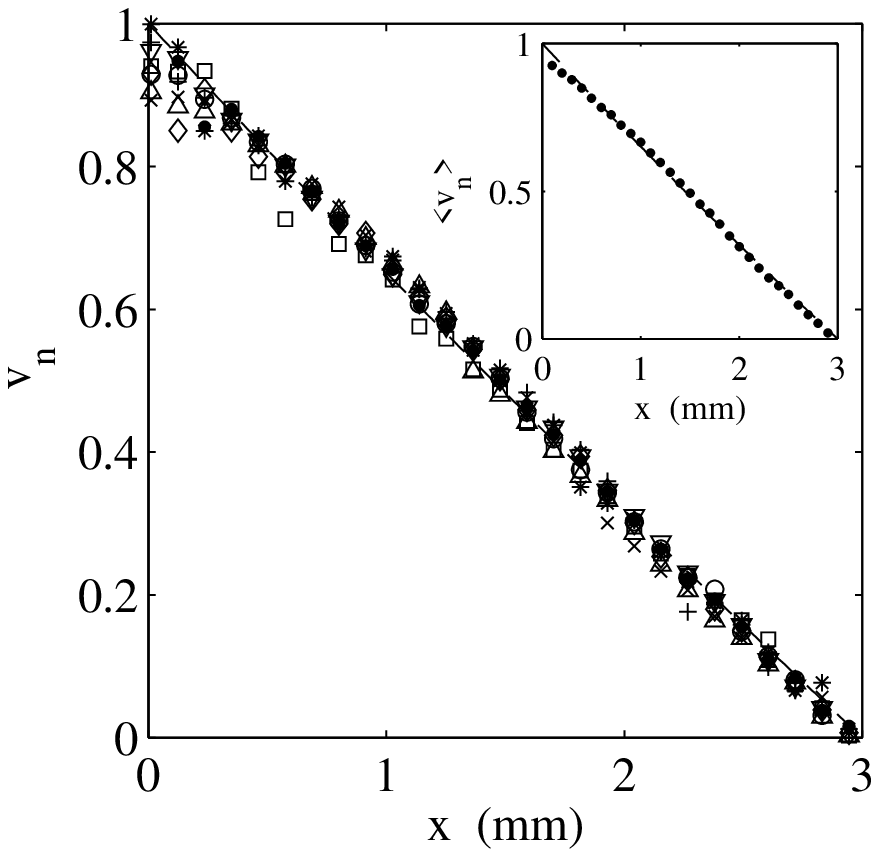}}
\end{center}
\caption{\label{emu20}Normalized velocity profiles measured in
the 20~\% emulsion for $\gammap=0.1$~s$^{-1}~(\triangledown)$, 
0.2~s$^{-1}~(\triangle)$, 0.4~s$^{-1}~(+)$, 1~s$^{-1}~(\times)$,
2~s$^{-1}~(\circ)$, 5~s$^{-1}~(\Box)$, 10~s$^{-1}~(\bullet)$,
15~s$^{-1}~(\diamond)$, and~20~s$^{-1}~(\ast)$.
Each profile is normalized according to
$v_n=(v-v_{\scriptscriptstyle 2})/(v_{\scriptscriptstyle 1}-v_{\scriptscriptstyle 2})$.
The dashed line
is the result expected for a Newtonian fluid.
Inset: the average of the previous
normalized profiles $<v_n>$ ($\bullet$)
compared to the Newtonian prediction (dashed line).}
\end{figure}

Figures~\ref{emu20_01_02}--\ref{emu20_10_15} present the velocity profiles measured
in the dilute emulsion of liquid fraction $\phi =20~\%$ 
for various imposed shear rates ranging from $0.1$~s$^{-1}$ to $15$~s$^{-1}$.
The velocity profiles expected for a Newtonian fluid are shown as dashed lines.
The velocities $v(x)$ have
been rescaled by the inner wall velocity $v_{\scriptscriptstyle 0}$.
In these figures, the upper left corner (at $x=0$) thus always corresponds to
the velocity of the inner wall moving at $v_{\scriptscriptstyle 0}$, and the lower right corner
(at $x=e=3$~mm) to the fixed outer wall ($v=0$).
First, one can notice that the experimental velocity is always 
different from zero at the stator and different from the rotor 
velocity at the rotor. This means that
significant slippage occurs and remains of the order of $10~\%$ of $v_{\scriptscriptstyle 0}$.
We define the slip velocity at the rotor $v_{s_1}=v_{\scriptscriptstyle 0}-v_{\scriptscriptstyle 1}$ as the difference between the 
rotor velocity $v_{\scriptscriptstyle 0}$ and the emulsion velocity $v_{\scriptscriptstyle 1}$ at the rotor 
and the slip velocity $v_{s_2}=v_{\scriptscriptstyle 2}$ at the stator
as the emulsion velocity $v_{\scriptscriptstyle 2}$ at the stator.

Moreover, all the velocity profiles measured in the dilute emulsion are linear.
To better illustrate this point, Fig.~\ref{emu20} presents the normalized velocity 
$v_n$ for various applied shear rates. We define $v_n$
experimentally by:
\begin{equation}
v_n(x)\,\widehat{=}\, \frac{v(x)-v_{2}}{v_{1}-v_{2}} \,.
\label{defvnorm}
\end{equation}
All normalized data coincide
and the average master curve $<v_n>$ is almost perfectly linear across the cell gap (see inset of
Fig.~\ref{emu20}).
These qualitative results show that, although subject to small but measurable wall slip,
the dilute emulsion flows like a Newtonian fluid in the range of shear rates under study.

\subsection{Concentrated emulsion ($\phi=75~\%$)}
\begin{figure}[htbp]
\begin{center}
\scalebox{0.45}{\includegraphics{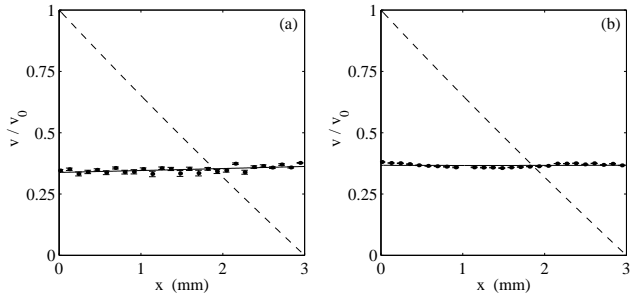}}
\end{center}
\caption{\label{emu75_01_02}Velocity profiles in
the 75~\% emulsion for (a) $\gammap=0.1$~s$^{-1}$
and (b) $\gammap=0.2$~s$^{-1}$.
The solid lines are linear fits to the data. The dashed curves
represent the velocity profiles expected for a Newtonian fluid.
}
\end{figure}

\begin{figure}[htbp]
\begin{center}
\scalebox{0.45}{\includegraphics{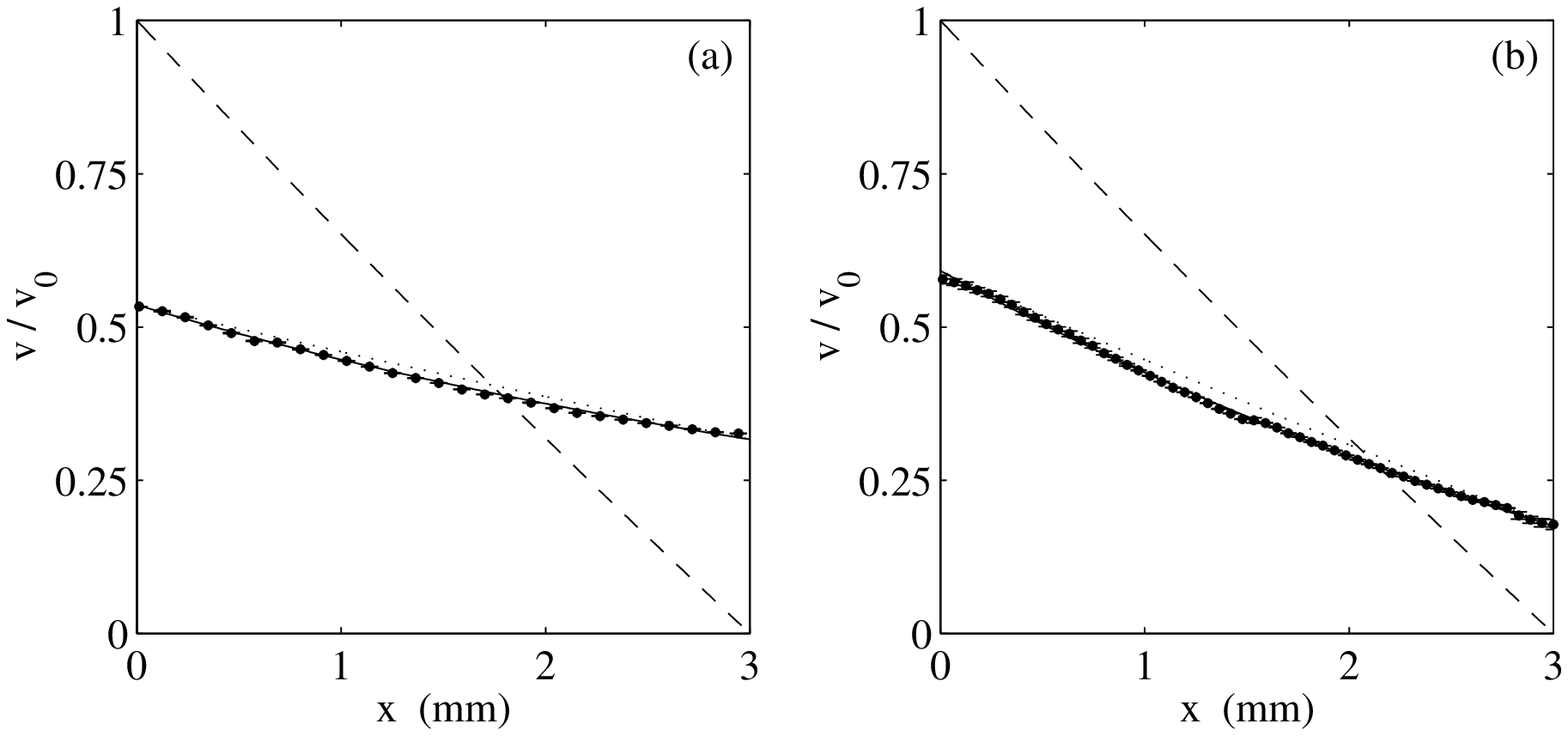}}
\end{center}
\caption{\label{emu75_04_1}Velocity profiles in
the 75~\% emulsion for (a) $\gammap=0.4$~s$^{-1}$
and (b) $\gammap=1$~s$^{-1}$.
The solid curves are the best fits to the data using Eq.~(\ref{vfit}) with $n=0.4$.
The dashed lines
represent the velocity profiles expected for a Newtonian fluid.
The dotted lines correspond to the case of a Newtonian fluid
that would slip at the cell walls with the same velocities
as the emulsion.}
\end{figure}

\begin{figure}[htbp]
\begin{center}
\scalebox{0.45}{\includegraphics{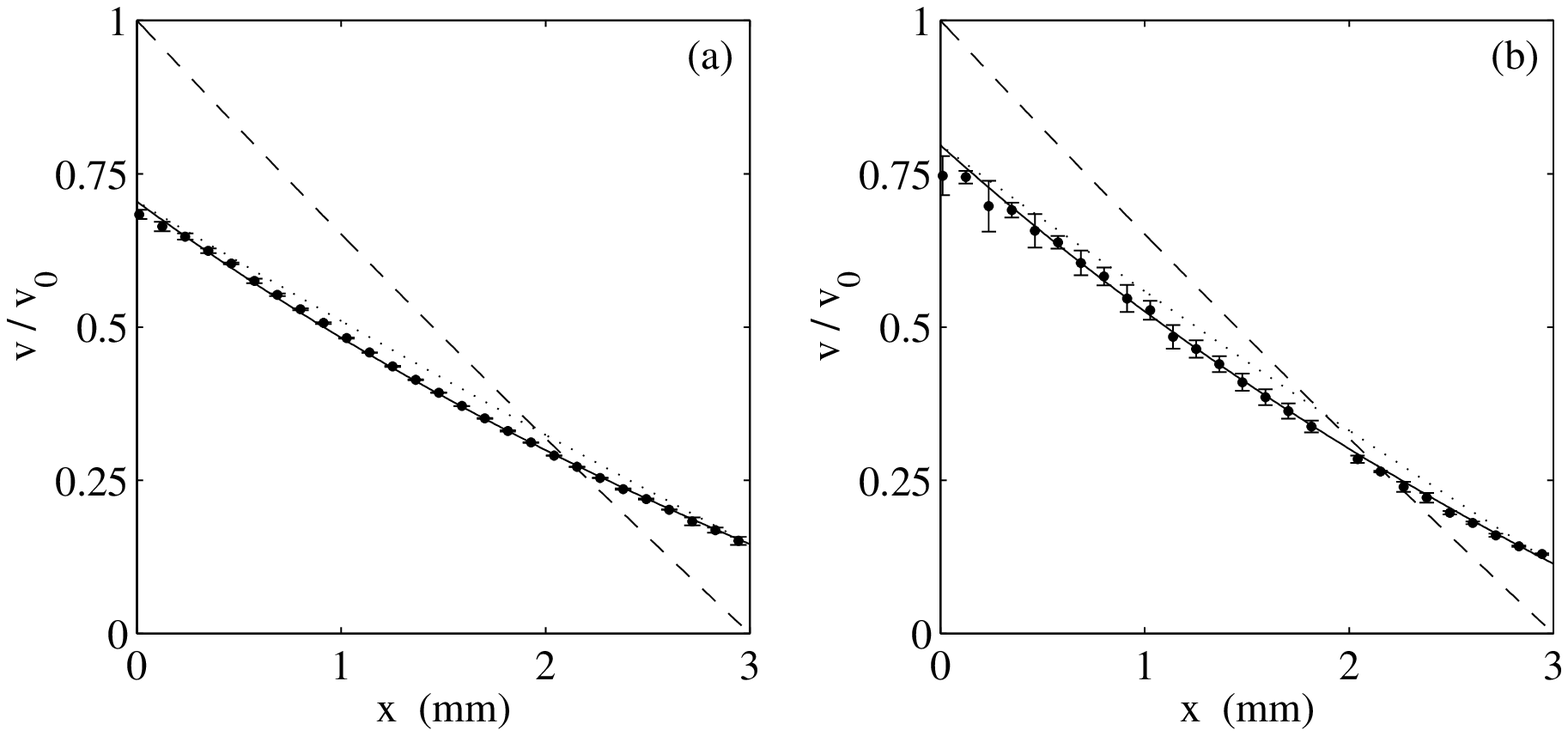}}
\end{center}
\caption{\label{emu75_2_5}Velocity profiles in
the 75~\% emulsion for (a) $\gammap=2$~s$^{-1}$
and (b) $\gammap=5$~s$^{-1}$.
The solid curves are the best fits to the data using Eq.~(\ref{vfit}) with $n=0.4$.
The dashed lines
represent the velocity profiles expected for a Newtonian fluid.
The dotted lines correspond to the case of a Newtonian fluid
that would slip at the cell walls with the same velocities
as the emulsion.}
\end{figure}

\begin{figure}[htbp]
\begin{center}
\scalebox{0.45}{\includegraphics{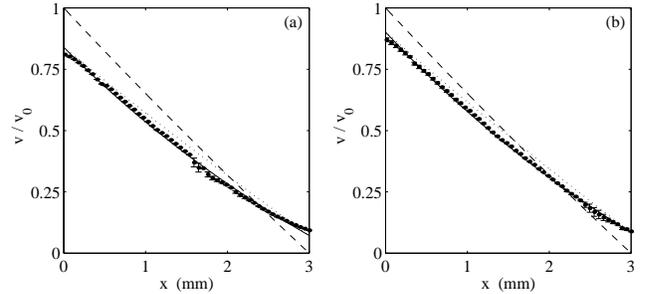}}
\end{center}
\caption{\label{emu75_10_15}Velocity profiles in
the 75~\% emulsion for (a) $\gammap=10$~s$^{-1}$
and (b) $\gammap=15$~s$^{-1}$.
The solid curves are the best fits to the data using Eq.~(\ref{vfit}) with $n=0.4$.
The dashed lines
represent the velocity profiles expected for a Newtonian fluid.
The dotted lines correspond to the case of a Newtonian fluid
that would slip at the cell walls with the same velocities
as the emulsion.}
\end{figure}

Figures~\ref{emu75_01_02}--\ref{emu75_10_15} present the
velocity profiles in the concentrated emulsion of liquid fraction $\phi = 75~\%$ 
for the previous range of imposed shear rates.
Two kinds of behaviours are encountered. In the case of low applied shear rates ($0.1$ and $0.2$~s$^{-1}$,
see Fig.~\ref{emu75_01_02}), 
the emulsion moves around in the gap as a solid body with angular velocity $\omega = 0.042$~rad.s$^{-1}$ for $\gammap =0.1$~s$^{-1}$
and $\omega = 0.084$~rad.s$^{-1}$ for $\gammap=0.2$~s$^{-1}$.
 In this regime, all the material actually experiencing shear is located inside thin
boundary layers at the two walls. 
These boundary layers cannot be resolved with our spatial resolution of about 100~$\mu$m.
When the applied shear rate is increased, 
the velocity profiles change. Wall slip remains very important but the emulsion no longer behaves as a solid
 body. Shear occurs not only in the boundary layers but also in the bulk material. 

\begin{figure}[htbp]
\begin{center}
\scalebox{0.6}{\includegraphics{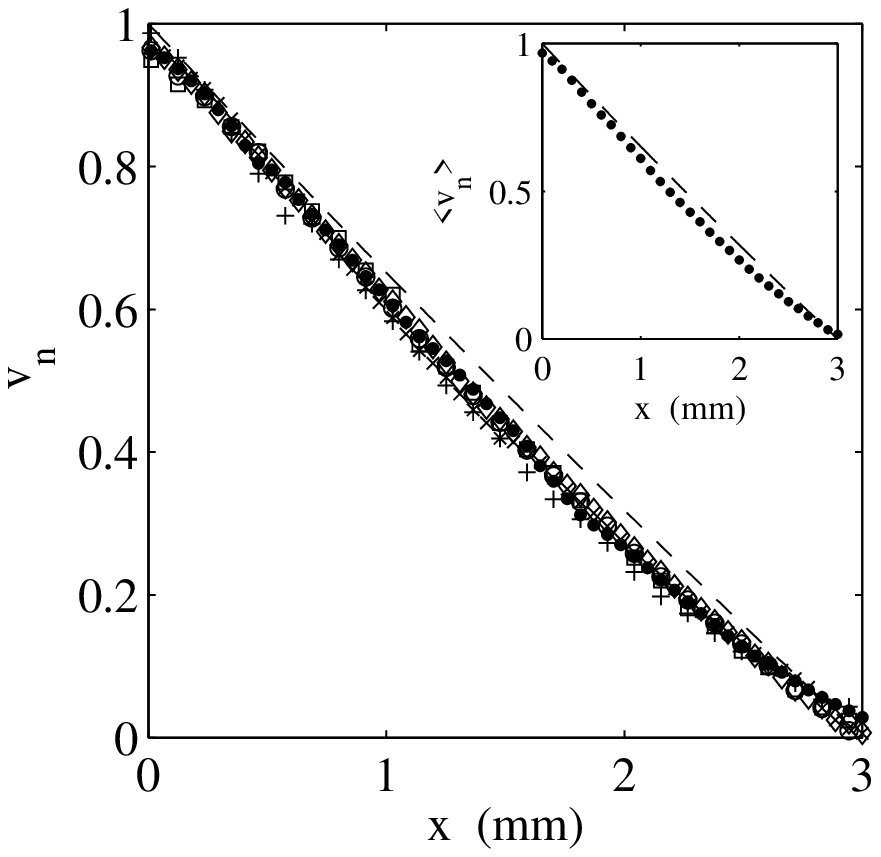}}
\end{center}
\caption{\label{emu75}Normalized velocity profiles measured in
the 75~\% emulsion for $\gammap=0.4$~s$^{-1}~(+)$, 1~s$^{-1}~(\times)$,
2~s$^{-1}~(\circ)$, 5~s$^{-1}~(\Box)$, 10~s$^{-1}~(\bullet)$,
and 15~s$^{-1}~(\diamond)$.
Each profile is normalized according to
$v_n=(v-v_{\scriptscriptstyle 2})/(v_{\scriptscriptstyle 1}-v_{\scriptscriptstyle 2})$.
The dashed line
is the result expected for a Newtonian fluid.
Inset: the average of the previous
normalized profiles $<v_n>$ ($\bullet$)
compared to the Newtonian prediction (dashed line).}
\end{figure}

The velocity profiles are no longer linear but rather curved (see Fig.~\ref{emu75_2_5} for instance). 
Moreover, as shown in Fig.~\ref{emu75}, the data for
the normalized velocity $v_n(x)$ still collapse rather well on a single curve
provided the applied shear rate is greater than $0.2$~s$^{-1}$. The curvature of the mean normalized
profile $<v_n>$ is clearly visible in the inset of Fig.~\ref{emu75}.
This result means that the non-Newtonian features of our concentrated emulsion are strong enough
to show up at a local level even in a gap as small as 3~mm.

In the next Section, we first discuss the evolution of the 
slip velocities and then analyze the features of individual velocity profiles in more depth.

\section{Local rheology of emulsions in a Couette flow}

\subsection{Wall-slip and wetting films}

For both the dilute and the concentrated emulsions, slippage clearly occurs. The simplest picture one
could imagine to describe this situation is the existence of two very thin films of
 continuous phase near the walls and a bulk material with the original concentration
 in between \cite{Princen:1985,Barnes:1995}. From our measurements, it is possible to access the thickness $h$ of the
 thin film where shear occurs. Indeed, assuming a slip velocity $v_s$ within
a thin liquid film of viscosity $\eta_f$, we may write:
$\sigma = \eta_{f} \gammap_s  = \eta_{f} v_{s}/h$, where $\gammap_s=v_s/h$ is the shear rate inside the film.
Using the subscript $i=1$ or 2 to denote the film at the rotor or at the stator respectively,
\begin{equation}
h_i=\frac{\eta_{f} v_{s_i}}{\sigma_i}\,\,\, \hbox{\rm for}\,\,\, i=1\,\,\, \hbox{\rm or}\,\,\, 2\,,
\label{hfilm}
\end{equation}
where $\sigma_i$ is the shear stress near wall number $i$. Note that, in the Couette geometry, the values of
$\sigma_i$ at the walls are linked to the global stress
$\sigma$ indicated by the rheometer according to:
\begin{equation}
\sigma_i=\frac{2R_j^2}{R_{\scriptscriptstyle 1}^2+R_{\scriptscriptstyle 2}^2}\,\sigma\,,
\label{e.sigmawall}
\end{equation}
where $j=2$ (resp. $j=1$) when $i=1$ (resp. $i=2$).
For each value of $\gammap$, the steady-state value of $\sigma$ imposed by the rheometer is recorded
and $\sigma_{\scriptscriptstyle 1}$ and $\sigma_{\scriptscriptstyle 2}$ are computed from Eq.~(\ref{e.sigmawall}).
In our case, this leads to significant deviations from $\sigma$: $\sigma_{\scriptscriptstyle 1}=1.13\,\sigma$ and
$\sigma_{\scriptscriptstyle 2}=0.87\,\sigma$.

\subsubsection{Dilute emulsion ($\phi=20~\%$)}
\begin{figure}[htbp]
\begin{center}
\scalebox{0.45}{\includegraphics{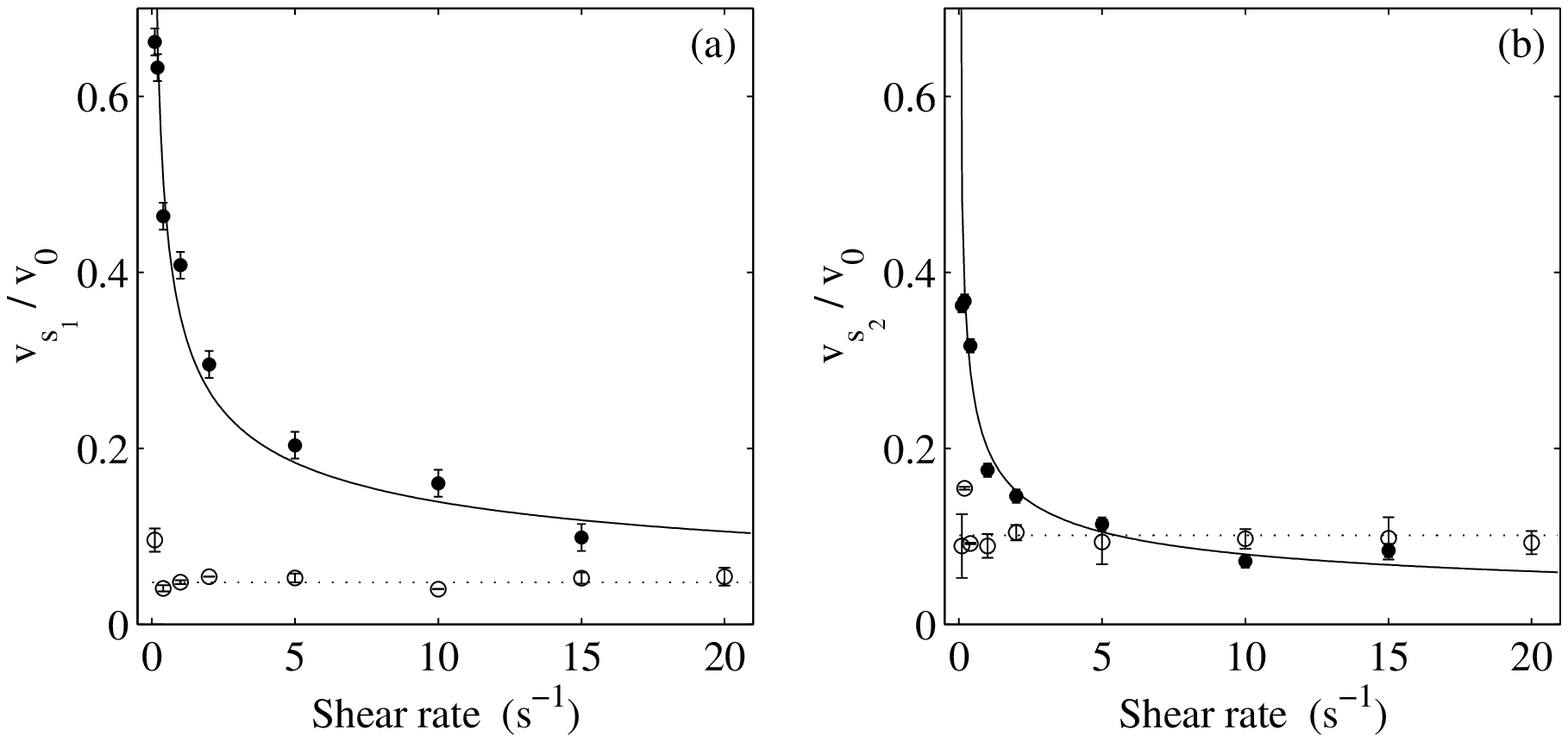}}
\end{center}
\caption{\label{gliss} Wall slip velocities for the 75~\% emulsion~($\bullet$)
and the 20~\% emulsion~($\circ$) vs. shear rate. (a) Normalized slip velocity $v_{s_1}/v_{\scriptscriptstyle 0}$
at the moving inner wall ($x=0$).
The solid line is the power law $0.35\,\gammap^{-0.4}$.
The dotted line shows the mean value $v_{s_1}=0.048\,v_{\scriptscriptstyle 0}$ for the
20~\% emulsion.
(b) Normalized slip velocity $v_{s_2}/v_{\scriptscriptstyle 0}$
at the fixed outer wall ($x=3$~mm).
The solid line is the power law $0.20\,\gammap^{-0.4}$.
The dotted line shows the mean value $v_{s_2}=0.10\,v_{\scriptscriptstyle 0}$ for the
20~\% emulsion.
}
\end{figure}

\begin{figure}[htbp]
\begin{center}
\scalebox{0.45}{\includegraphics{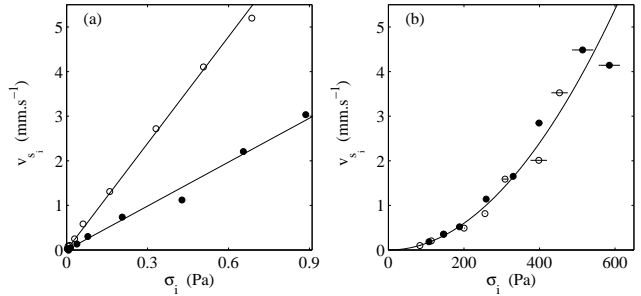}}
\end{center}
\caption{\label{figvit} Slip velocities vs. shear stress: $v_{s_1}$ vs. $\sigma_{\scriptscriptstyle 1}$ at the rotor ($\bullet$) and $v_{s_2}$ vs. $\sigma_{\scriptscriptstyle 2}$
at the stator ($\circ$). 
(a) For the 20~\% emulsion. The straight lines correspond to $v_{\scriptscriptstyle 1}=33\sigma_{\scriptscriptstyle 1}$ and to $v_{\scriptscriptstyle 2}=8.0\sigma_{\scriptscriptstyle 2}$.
(b) For the 75~\% emulsion. The solid line is $v_i=1.5.10^{-5}\,\sigma_i^2$.
}
\end{figure}

In the dilute system, the normalized slip velocities at the rotor $v_{s_1}/v_{\scriptscriptstyle 0}$
and at the stator $v_{s_2}/v_{\scriptscriptstyle 0}$ are independent of the applied shear rate as
can be checked on Fig.~\ref{gliss}. This means that $v_{s_i}\sim\gammap$ and since the dilute
emulsion has a Newtonian behaviour, one expects $v_{s_i}\sim\sigma_i$. This scaling law is
indeed observed on Fig.~\ref{figvit}(a).

\begin{figure}[htbp]
\begin{center}
\scalebox{0.45}{\includegraphics{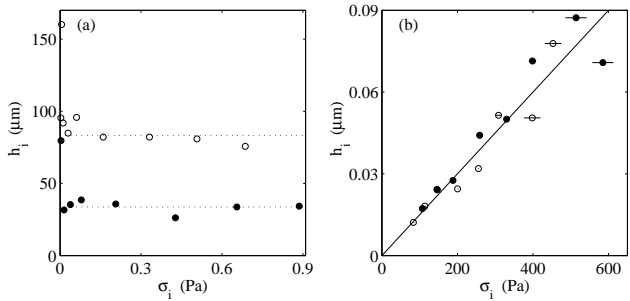}}
\end{center}
\caption{\label{figfilm} Thickness of the sliding
layers vs. shear stress: $h_{\scriptscriptstyle 1}$ vs. $\sigma_{\scriptscriptstyle 1}$ at the rotor ($\bullet$) and $h_{\scriptscriptstyle 2}$ vs. $\sigma_{\scriptscriptstyle 2}$
at the stator ($\circ$). 
(a) For the 20~\% emulsion. The dashed lines are $h_{\scriptscriptstyle 1}=34~\mu$m and $h_{\scriptscriptstyle 2}=83~\mu$m.
(b) For the 75~\% emulsion. The solid line corresponds to $h_i=1.5.10^{-4}\,\sigma_i$.
}
\end{figure}

Assuming that the films contain only water and glycerol, one has
$\eta_{f} = 0.01$~Pa.s. The thicknesses of the films $h_{\scriptscriptstyle 1}$ and $h_{\scriptscriptstyle 2}$ calculated from Eq.~(\ref{hfilm})
are shown as functions of $\sigma_{\scriptscriptstyle 1}$ and $\sigma_{\scriptscriptstyle 2}$ in Fig.~\ref{figfilm}(a).
One gets $h_{\scriptscriptstyle 1}\approx 30~\mu$m at the rotor and $h_{\scriptscriptstyle 2}\approx 80~\mu$m 
at the stator independent of the shear stress, which is again consistent with a Newtonian behaviour.
Remarkably, these values are much higher than the diameter of the droplets ($2~\mu$m), meaning that depletion
in the dilute emulsion extends very far from the walls. Various explanations may be proposed for such
a large depletion effect.

First, depletion on such long distances might be induced by cumulative effects of
inertial effect and of cross-stream migration. 
Indeed, Goldsmith and Mason reported rapid migration from walls of
neutrally buoyant and deformable droplets in pipe and Couette flows \cite{Goldsmith:1962}. 
The authors suggest that such a  migration could be initiated by
the presence of the wall and of non-uniform velocity gradients.
A net force acting to push the drop to a lower gradient is created. 
 
Second, it is important to note that the viscosity of the emulsion varies
dramatically with oil concentration. 
For example, the viscosity of the 20~\% emulsion indicated by the rheometer
is 0.037~Pa.s. When
diluted to a volume fraction $\phi = 10~\%$, the viscosity decreases
to 0.017~Pa.s. Finally, at $\phi=5~\%$, the viscosity is equal to 0.011~Pa.s,
very close to that of the continuous phase.
This means that the lubricating layers could also be viewed as emulsion films
with a concentration gradient, for example from  $\phi=20~\%$ in the bulk to $\phi<5~\%$ near the walls, rather than
a film of pure continuous phase.
This would reduce greatly the magnitude of the depletion and could extend its effect over large distances.

Finally, the thicknesses of the films are different at the rotor and at the stator. 
Slippage occurs preferentially at the stator (about $10~\%$ of $v_{\scriptscriptstyle 0}$) and remains about two times
smaller at the moving inner wall (about $5~\%$ of $v_{\scriptscriptstyle 0}$). Again,
since the oil density is slightly below that of the water--glycerol mixture,
this might be due to inertial effects
that lead to different compositions of the two sliding layers.

\subsubsection{Concentrated emulsion ($\phi=75~\%$)}

In the concentrated system, the thicknesses of the lubricating layers are much smaller
because for a given shear rate, the stress in the emulsion is much larger. 
It is clear that cross-migration and inertial effects are completely suppressed in this case.
As shown in Fig.~\ref{figfilm}(b),
the thickness of the films calculated from Eq.~(\ref{hfilm}) increases from $h_{\scriptscriptstyle 1}\approx10$~nm
at low shear stress to $h_{\scriptscriptstyle 1}\approx90$~nm at high shear stress.
These thicknesses are of the order of magnitude of that
of the thin liquid films of continuous phase that lie between the droplets.
Such results are in qualitative agreement with
previous measurements performed by Princen
using direct visualization of a stripe painted on the emulsion surface \cite{Princen:1985}.

Several reasons may be given for the increase of 
$h$ as a function of shear stress. (i) $h$ may increase as a result of the cell 
walls dragging bulk continuous phase from the pockets between the drops inside the films. (ii) The viscosity of 
the wetting film may decrease due to an increased mobility of the drop/film interface as the 
stress increases.

Like in the dilute emulsion, the phenomenon of wall slip may not seem symmetrical
when one plots $v_{s_i}/v_{\scriptscriptstyle 0}$ vs. $\gammap$ (see Fig.~\ref{gliss}).
This time, for a given shear rate, the slip velocity
is about 50~\% larger at the rotor than at the stator.
However when plotted against the local shear stresses at the walls,
the data $v_{s_1}(\sigma_{\scriptscriptstyle 1})$ and  $v_{s_2}(\sigma_{\scriptscriptstyle 2})$ collapse on each other as can be seen on Fig.~\ref{figvit}(b).
One can thus interpret the difference between $v_{s_1}(\gammap)$ and $v_{s_2}(\gammap)$
as a consequence of the difference
between $\sigma_{\scriptscriptstyle 1}$ and $\sigma_{\scriptscriptstyle 2}$ due to the inhomogeneity of the 
shear stress in the gap. Indeed, in our Couette cell,
the shear stresses at the rotor and at the stator
differ by about 25~\%. One then recovers a 50~\% difference in the slip velocities
provided $v_{s_i}\sim\sigma_i^2$. This scaling behaviour is evidenced in Fig.~\ref{figvit}(b) where
a parabola is seen to fit nicely all the $v_s(\sigma)$ data. A direct consequence
of Eq.~(\ref{hfilm}) is that
$h_{\scriptscriptstyle 1}(\sigma_{\scriptscriptstyle 1})=h_{\scriptscriptstyle 2}(\sigma_{\scriptscriptstyle 2})$
and that $h(\sigma)\sim\sigma$ (see Fig.~\ref{figfilm}(b)).

\subsubsection{Discussion on usual corrections for wall slip}

The previous results are important because they provide a confirmation
that the corrections for wall slip proposed by Yoshimura and Prud'homme \cite{Yoshimura:1988a} or by
Kilja\'nski \cite{Kiljanski:1989} based on those introduced by Mooney \cite{Mooney:1931}
are relevant for concentrated emulsions.
Indeed, these authors assumed (i)~that the slip velocity and
the thickness of sliding layers are functions of the shear stress only and
(ii)~that the slip velocities at the rotor and at the stator are the
same functions of $\sigma$.
Then, from rheological experiments performed in different geometries (by varying
the height \cite{Yoshimura:1988a} or the gap \cite{Kiljanski:1989} of the Couette cell),
corrections for wall slip are computed, that yield an indirect
estimate of $v_s(\sigma)$.
The present results based on {\it direct} measurements of $v_s$ at the
two walls show that
$v_{s_1}(\sigma_{\scriptscriptstyle 1})=v_{s_2}(\sigma_{\scriptscriptstyle 2})$ and reveal a quadratic behaviour
of $v_s$ vs. $\sigma$ consistent with Refs.~\cite{Yoshimura:1988a,Kiljanski:1989}.

However, the indirect corrections of Refs.~\cite{Yoshimura:1988a,Kiljanski:1989}
are clearly not valid for our dilute emulsion since Fig.~\ref{figvit} shows that
$v_{s_1}(\sigma_{\scriptscriptstyle 1})\neq v_{s_2}(\sigma_{\scriptscriptstyle 2})$. In this case, local measurements are
required to estimate accurately the influence of wall slip on rheological data.

\subsection{Local rheology vs. global rheology}

\subsubsection{Effective shear rate}

\begin{figure}[htbp]
\begin{center}
\scalebox{0.6}{\includegraphics{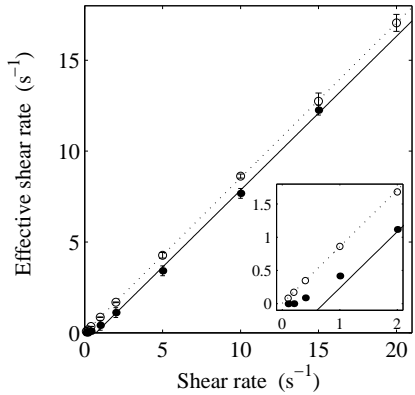}}
\end{center}
\caption{\label{cis_n} Effective shear rate in the
75~\% emulsion~($\bullet$)
and in the 20~\% emulsion~($\circ$) vs. imposed shear rate.
The solid line is the best linear fit of the 75~\% data for
$\gammap\geq 1$~s$^{-1}$:
$\gammap_{\hbox{\rm \scriptsize eff}}=0.85\,\gammap-0.61$.
The dotted line is the best linear fit of the 20~\% data:
$\gammap_{\hbox{\rm \scriptsize eff}}=0.85\,\gammap+0.005$.
Inset: blow-up of the low shear regime.
}
\end{figure}

Using the slip velocities estimated from the velocity profiles,
it is easy to define an effective shear rate in the
emulsion by $\gammapeff=(\max(v)-\min(v))/e=(v_{\scriptscriptstyle 1}-v_{\scriptscriptstyle 2})/e$.
However, in order to take into account the stress inhomogeneity in the Couette cell and
thus get a quantitative comparison between this effective shear rate and the global
shear rate $\gammap$ given by the rheometer, we have to define $\gammapeff$ consistently
with Eqs.~(\ref{e.gammarheo}) and (\ref{e.gammacorr}) by:
\begin{equation}
\gammapeff=\frac{R_{\scriptscriptstyle 1}^2+R_{\scriptscriptstyle 2}^2}{R_{\scriptscriptstyle 1}(R_{\scriptscriptstyle 1}+R_{\scriptscriptstyle 2})}\,\frac{v_{\scriptscriptstyle 1}-v_{\scriptscriptstyle 2}}{e}\,.
\label{e.effectiveshear}
\end{equation}
In Fig.~\ref{cis_n}, $\gammapeff$ is plotted against the imposed shear rate $\gammap$ for the two emulsions.
Surprisingly both data sets are very well fitted by straight lines with the exact same slope 0.85
(provided $\gammap\ge 1$~s$^{-1}$ for the concentrated emulsion, see inset of Fig.~\ref{cis_n}).
For the dilute Newtonian emulsion, $\gammapeff\approx0.85\gammap$ only confirms that the emulsion undergoes a
constant wall slip of about 15~\% independent of $\gammap$.
This implies that due to wall slip,
the value of the viscosity $\eta=\sigma/\gammap$ given by the rheometer is underestimated
by about 15~\% so that the actual viscosity is $\eta\approx 0.043$~Pa.s instead of 0.037~Pa.s.
The fact that $\gammapeff\approx0.85\gammap+\hbox{\rm cst}$ also
for the 75~\% emulsion seems to indicate that this
value of 0.85 is somehow characteristic of the oil droplets (radius, surface tension, etc.).
Further studies should focus on this specific point.

\subsubsection{Global flow curve(s)}

In the particular case of the concentrated system, it is interesting to compare the classical
rheological data $\sigma(\gammap)$ to the ``effective flow curve''
$\sigma(\gammapeff)$ that can be estimated from
the local velocity measurements. 

Classical steady-state flow measurements were carried out using both a cone-and-plate geometry with 
striated surfaces (cone angle $2^\circ$ and radius 20~mm)
and the Couette cell with smooth walls used throughout this study.
The emulsion sample was the same as the one
used for the velocity profile measurements. Striated surfaces are supposed to minimize
wall slip whereas the measurements obtained with the smooth walls lead to the values of $\sigma$
and $\gammap$ used up to now.

\begin{figure}[htbp]
\begin{center}
\scalebox{0.6}{\includegraphics{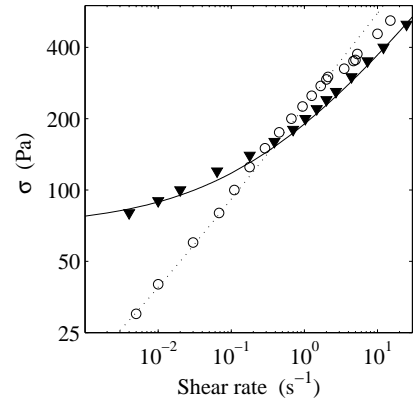}}
\end{center}
\caption{\label{rheol75glob}Rheological flow curves $\sigma$ vs. $\gammap$ for the 75~\%
emulsion measured in a striated cone-and-plate geometry ($\blacktriangledown$)
and in the Plexiglas Couette cell with smooth walls and $e=3$~mm ($\circ$).
The solid line is $\sigma=70+120\,\gammap^{0.4}$ and the equation of the dotted line is
$\sigma=4+220\,\gammap^{0.4}$.}
\end{figure}

A constant stress is applied on the sample for ten minutes and the steady-state shear rate is recorded.
The flow curves are obtained point by point from the lowest stress upwards.
We checked that measurements from the highest stress
downwards give the same results. The two steady-state flow curves $\sigma(\gammap)$ obtained
for the concentrated emulsion are compared on logarithmic scales in Fig.~\ref{rheol75glob}. 
The two curves are clearly different. 
If one considers only the measurements obtained with the Couette geometry, 
it is not obvious whether the emulsion has a yield stress or not.
Indeed, the yield stress inferred from this data set is very small: $\seuil\approx 4$~Pa.
On the other hand, measurements obtained with the striated cone-and-plate geometry clearly
present a yield stress
of about 70~Pa.

However, even when using striated surfaces, global rheological measurements may still
be affected by small
slippage or by fracture behaviour at shear rates close to the yield point \cite{Mason:1996a}.
The determination of a precise value for the yield stress is thus difficult
with standard rheological tools. A way around the problem
is to measure the local velocity and plot the shear stress as a function of the effective shear rate 
$\gammapeff$ in
the sample.

\subsubsection{Local flow curve}

\begin{figure}[htbp]
\begin{center}
\scalebox{0.6}{\includegraphics{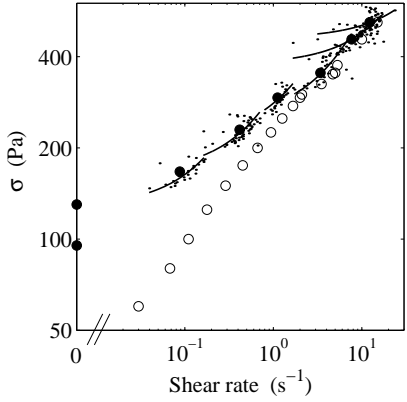}}
\end{center}
\caption{\label{rheol75loc}The ``effective flow curve'' $\sigma$ vs. $\gammapeff$
($\bullet$) and the ``local flow curve'' $\sigma(x)$ vs. $\gammap(x)$ (dots)
for the 75~\%
emulsion compared to the rheological data obtained in the Plexiglas Couette cell ($\circ$).
The solid segments are linear fits of the local data $\sigma(x)$ vs. $\gammap(x)$.}
\end{figure}

As seen in Fig.~\ref{rheol75loc} where the ``effective flow curve''
$\sigma(\gammapeff)$ is compared
to the global rheological data obtained in the Couette cell, $\sigma(\gammapeff)$
yields yet another flow curve, different from the curves of Fig.~\ref{rheol75glob}.
Now, the two points where the emulsion does not flow lie on the axis $\gammap=0$ and 
the data corresponding to $\gammap\ge 0.4$~s$^{-1}$ seem to point to a higher value of the
yield stress around 90~Pa. As we shall see below, the present data do not allow
us to draw definite conclusions about the value of the yield stress nor even about the
existence of a simple analytic flow curve for the concentrated emulsion.

Since the velocity profiles are curved, one may argue that plotting the effective
shear rate $\gammapeff$
against the shear stress given by the rheometer is not such an accurate estimate of
the flow curve.
However, using local measurements, one can go one step further and estimate the
local shear rate $\gammap(x)$ by differentiating the velocity profile $v(x)$ according to:
\begin{equation}
\gammap(x)=-(R_{\scriptscriptstyle 1}+x)\,\frac{\partial}{\partial x}\left(\frac{v(x)}{R_{\scriptscriptstyle 1}+x}\right)\, .
\label{e.localshearrate}
\end{equation}
This was done directly on the raw velocity data with a first-order approximation
for the derivative: $\partial v/\partial x\approx\delta v/\delta x=(v(x+\delta x)-v(x))/\delta x$.
The minus sign in Eq.~(\ref{e.localshearrate}) accounts for the orientation of the $x$ axis along
decreasing velocities.

Moreover, if {\it the flow is stationary} and {\it the sample does not present any fracture}, then
$\nabla\sigma=0$ and the local shear stress reads:
\begin{equation}
\sigma(x) = \sigma_{\scriptscriptstyle 1}\left(\frac{R_{\scriptscriptstyle 1}}{R_{\scriptscriptstyle 1} + x}\right)^2\, ,
\label{sigmax}
\end{equation}  
where $R_{\scriptscriptstyle 1}$ is the radius of the rotor and $\sigma_{\scriptscriptstyle 1}=\sigma(x=0)$
the shear stress applied by the rheometer
on the rotor. Note that this equation was already implicitely used at the walls when
writing Eq.~(\ref{e.sigmawall}). For each velocity
profile and for each value of $x$, the point $(\gammap(x),\sigma(x))$ obtained
by Eqs.~(\ref{e.localshearrate}) and (\ref{sigmax}) is plotted in Fig.~\ref{rheol75loc}.

We check that the data $(\gammap(x),\sigma(x))$ are scattered around the effective flow curve
$\sigma(\gammapeff)$. The dispersion seen in this ``local flow curve'' $(\gammap(x),\sigma(x))$
mainly arises from the experimental estimation of the derivative in Eq.~(\ref{e.localshearrate}).
For each shear rate, each cluster of points $(\gammap(x),\sigma(x))$ was fitted by a straight line.
The resulting segments are plotted in logarithmic scales in Fig.~\ref{rheol75loc}.
It can be seen that $(\gammapeff,\sigma)$ always fall close to these first order approximations
of the local flow curve. Note that due to the inhomogeneity of the stress in the Couette cell,
the shear rates explored locally continuously cover the whole range 0.03~s$^{-1}$--20~s$^{-1}$.

However, a jump in the local behaviour can be clearly identified
between $\gammap=1$~s$^{-1}$ and $\gammap=5$~s$^{-1}$ and the local flow curve does not appear to
be continuous at larger shear rates. Thus, within the precision of our data,
it looks as though the local flow behaviour differs from one imposed shear rate to another, at least
for $\gammap\ge1$~s$^{-1}$.
These results indicate that the very notion of a simple analytic flow curve may become questionable at
high shear rates for our concentrated
emulsion. In the next paragraph, we try to make this point clearer by focussing on the precise shape
of the velocity profiles.

\subsection{Trying to account for the curvature of the velocity profiles}

\subsubsection{Testing the flow behaviour $\sigma=A\gammap^n$}

Using MRI, Coussot {\it et al.} \cite{Coussot:2002} 
found that above some critical stress $\sigma_c$,
the velocity profiles of various flowing concentrated materials
were well described by a local power-law behaviour:
\begin{eqnarray}
\gammap(x)&=&0\,\,\, \hbox{\rm for} \,\,\,\sigma < \sigma_c \, ,
\label{jamming} \\
\sigma(x)&=&A\, \gammap(x)^n \,\,\, \hbox{\rm for} \,\,\,\sigma \ge \sigma_c\, .
\label{flowcurve}
\end{eqnarray}
Those two equations describe a flow behaviour with two separate branches.
Equation~({\ref{jamming}) represents the jammed state and holds below the yield point $\sigma_c$.
Equation~({\ref{flowcurve}) corresponds to a power-law fluid.
The case $n=1$ and $\sigma_c=0$ corresponds to the case of a Newtonian fluid
($A$ is then the usual viscosity $\eta$). $n<1$ (resp.
$n>1$) means that the fluid is shear-thinning (resp. shear-thickening).
This exponent $n$ controls the curvature of the velocity profiles $v(x)$.
In light of the results of Ref.~\cite{Coussot:2002}, we wanted to test whether Eq.~(\ref{flowcurve})
was able to account for the shape
of our velocity profiles when the emulsion flows {\it i.e.} for
$\gammap > 0.2$~s$^{-1}$.

\paragraph{Theoretical velocity profiles.}
Using Eqs.~(\ref{e.localshearrate}), (\ref{sigmax}) and (\ref{flowcurve}),
it is straightforward to calculate the flow profile above the yield point:
\begin{align}
v(x) &= v_{\scriptscriptstyle 2}\,\frac{R_{\scriptscriptstyle 1}+x}{R_{\scriptscriptstyle 2}}\, +\notag\\ &
(R_{\scriptscriptstyle 1}+x)\,\frac{n}{2} \left( \frac{\sigma_{\scriptscriptstyle 1}}{A}
\left(\frac{R_{\scriptscriptstyle 1}}{R_{\scriptscriptstyle 2}}\right)^2 \right)^{1/n} 
\left( \left(\frac{R_{\scriptscriptstyle 2}}{R_{\scriptscriptstyle 1}+x}\right)^{2/n} -1\right) \, , 
\label{profiltheo}
\end{align}
and the velocities of the emulsion at the two walls are linked by:
\begin{align}
v_{\scriptscriptstyle 1}=v_{\scriptscriptstyle 2}\,\frac{R_{\scriptscriptstyle 1}}{R_{\scriptscriptstyle 2}} + R_{\scriptscriptstyle 1}\,\frac{n}{2} \left( \frac{\sigma_{\scriptscriptstyle 1}}{A} \left(\frac{R_{\scriptscriptstyle 1}}{R_{\scriptscriptstyle 2}}\right)^2 \right)^{1/n} 
\left( \left(\frac{R_{\scriptscriptstyle 2}}{R_{\scriptscriptstyle 1}}\right)^{2/n} -1\right) \, . 
\label{lien}
\end{align}
As before, $\sigma_{\scriptscriptstyle 1}$ denotes the shear stress at the moving inner wall:
$\sigma_{\scriptscriptstyle 1}=\sigma(x=0)$. Using Eqs.~(\ref{profiltheo}), (\ref{lien}), and
the definition of Eq.~(\ref{defvnorm}), the normalized flow profile is given by: 
\begin{align}
v_n(x) \approx \left(1+\frac{x}{R_{\scriptscriptstyle 1}}\right)\,
\frac{ \left(\frac{R_{\scriptscriptstyle 2}}{R_{\scriptscriptstyle 1}+x}\right)^{2/n} -1 }{\left(\frac{R_{\scriptscriptstyle 2}}{R_{\scriptscriptstyle 1}}\right)^{2/n} -1} \, .
\label{vnorm}
\end{align}
The previous approximation results from the assumption that $v_{\scriptscriptstyle 1}/R_{\scriptscriptstyle 2}\ll \gammap(x)$ everywhere in the cell gap.
Since $v_{\scriptscriptstyle 1}\lesssim v_{\scriptscriptstyle 0}$,
$\gammap(x)\lesssim v_{\scriptscriptstyle 0}/e$, and
$R_{\scriptscriptstyle 2}\simeq 10\,e$, Eq.~(\ref{vnorm}) holds with an accuracy
that is always better than the experimental uncertainty of about 5~\%.

\paragraph{Experimental determination of the exponent $n$.}
Equation~(\ref{vnorm}) shows that in the framework of Eq.~(\ref{flowcurve}), focussing on $v_n(x)$
allows us to remove the contributions of both the slip velocity $v_{\scriptscriptstyle 2}$ and the prefactor $A$
to the velocity profiles, so that the only free parameter is $n$.
Figure \ref{nonlin}(a) presents theoretical profiles obtained with Eq.~(\ref{vnorm}) for
$R_{\scriptscriptstyle 1}=22$~mm, $R_{\scriptscriptstyle 2}=25$~mm, and values of $n$
ranging from 0.2 to 1. The inset shows that velocity profiles can hardly be distinguished from one another
when $0.6\le n\le 1$. In Fig.~\ref{nonlin}(b), we plotted the master curve $<v_n(x)>$ found
for the concentrated emulsion (inset of Fig.~\ref{emu75}) together
with theoretical profiles for $n=0.3$, 0.4, and 0.5 given by Eq.~(\ref{vnorm}). The experimental data
and their error bars fall between the theoretical profiles with $n=0.3$ and $n=0.5$ so that
we have $n=0.4 \pm 0.1$ 
in the case of the concentrated emulsion. This value of $n$
is compatible with both the model of Princen \cite{Princen:1989} that predicts 
an exponent $n=1/2$ and that of Berthier {\it et al.} \cite{Berthier:2001}
which yields $n=1/3$ close to the glass transition in soft materials.

\begin{figure}[htbp]
\begin{center}
\scalebox{0.45}{\includegraphics{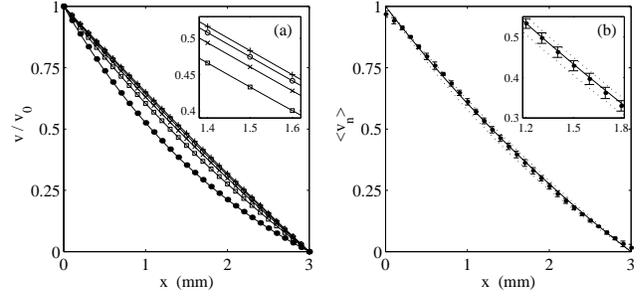}}
\end{center}
\caption{\label{nonlin}(a) Theoretical velocity profiles
obtained with Eq.~(\ref{vnorm}) for $n=0.2$ ($\bullet$), 0.4 ($\Box$), 0.6
($\times$), 0.8 ($\circ$), and 1 ($+$). Inset: blow-up of the middle
of the gap. (b) Average normalized profiles for the 75~\%
emulsion $<v_n>$ ($\bullet$)
and the non-Newtonian predictions of Eq.~(\ref{vnorm})
with $n=0.3$ (lower dotted line),
$n=0.4$ (solide line), and $n=0.5$ (upper dotted line).
Inset: blow-up of the middle of the gap.
}
\end{figure}

When performed on the dilute emulsion, the same analysis yields
$n=1 \pm 0.1$. In this case, the very small curvature of $v_n(x)$ due to the
variation of $\sigma(x)$ across the gap
cannot be detected and the Newtonian profile is undistinguishable from a straight line, at
least in a Couette cell
with $R_{\scriptscriptstyle 1}=22$~mm and $e=3$~mm. On the other hand, when the fluid is shear-thinning enough, say when $n<0.6$, the effect
of the shear stress non-uniformity is enhanced and curvature becomes quantitatively measurable even
in a Couette cell with a rather small gap. All the velocity profiles obtained with the 
concentrated emulsion are consistent with a single value of $n=0.4\pm 0.1$
and more experiments, perhaps in wider gaps, could help discriminate
between $n=1/3$ and $n=1/2$. However, for the model to be fully applicable, one still
has to check that Eq.~(\ref{flowcurve}) holds for a single value of the parameter $A$
over the whole range of investigated shear rates.

\paragraph{Experimental determination of the parameter $A$ and failure of the model.}
Once $n$ is known, it is possible to estimate the prefactor $A$ from our experimental data.
More precisely, the data in Figs.~\ref{emu20_01_02}--\ref{emu20_10_15} and \ref{emu75_04_1}--\ref{emu75_10_15}
were fitted using a least-square algorithm by:
\begin{equation}
v(x) = B\,\frac{R_{\scriptscriptstyle 1}+x}{R_{\scriptscriptstyle 2}} + C\,(R_{\scriptscriptstyle 1}+x) \left( \left(\frac{R_{\scriptscriptstyle 2}}{R_{\scriptscriptstyle 1}+x}\right)^{2/n} -1\right) \, , 
\label{vfit}
\end{equation}
where $B$ and $C$ are free parameters and $n$ was fixed to $n=1$ for the 20~\% emulsion and $n=0.4$ for
the 75~\% emulsion. The best fits are plotted in solid lines in the corresponding figures. In all cases,
those fits describe the experimental data very closely.
Note that for the concentrated emulsion at $\gammap=0.1$~s$^{-1}$
and $0.2$~s$^{-1}$ (Fig.~\ref{emu75_01_02}), the data were fitted by straight lines since the emulsion
undergoes solid-body rotation in this low shear regime.

Moreover, if Eq.~(\ref{profiltheo}) holds, one expects:
\begin{equation}
B=v_{\scriptscriptstyle 2}\,\,\,\hbox{\rm and}\,\,\,
C=\frac{n}{2} \left( \frac{\sigma_{\scriptscriptstyle 1}}{A} \left(\frac{R_{\scriptscriptstyle 1}}{R_{\scriptscriptstyle 2}}\right)^2 \right)^{1/n}\, .
\end{equation}
Hence, if the power-law behaviour of Eq.~(\ref{flowcurve}) is verified, one should have:
\begin{equation}
D\,\widehat{=}\, {\left(\frac{R_{\scriptscriptstyle 2}}{R_{\scriptscriptstyle 1}}\right)}^2 {\left(\frac{2C}{n}\right)}^n 
=\frac{\sigma_{\scriptscriptstyle 1}}{A} \, .
\label{e.D}
\end{equation}

Thus, an important self-consistency check of the model
can be performed by plotting $D$ against the shear stress at the rotor
$\sigma_{\scriptscriptstyle 1}$.
As shown in Fig.~\ref{prefact}(a), the data for the dilute emulsion are very well fitted
by $D=\sigma_{\scriptscriptstyle 1}/A$ with $A\approx 0.042$. This is consistent with the Newtonian behaviour 
of the 20~\% emulsion and with the value found for its viscosity from rheological 
measurements when corrected for wall slip ($\eta=0.043$~Pa.s).

\begin{figure}[htbp]
\begin{center}
\scalebox{0.45}{\includegraphics{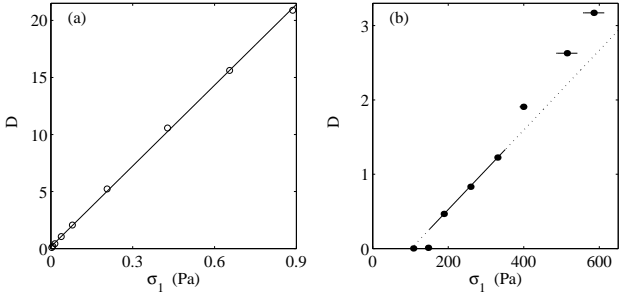}}
\end{center}
\caption{\label{prefact}The parameter $D$ vs. $\sigma$ (see text). (a)
For the 20~\% emulsion ($\circ$), the best linear fit
yields $A=0.042$.
(b) For the 75~\% emulsion~($\bullet$), the best linear fit for $150<\sigma_{\scriptscriptstyle 1}<350$~Pa
({\it i.e.} $0.4\le\gammap\le 1$~s$^{-1}$) 
does not go to zero when $\sigma_{\scriptscriptstyle 1}\to 0$.}
\end{figure}

As for the concentrated emulsion, the behaviour is quite different. Figure~\ref{prefact}(b) shows
that the parameter $D$ no longer behaves linearly with $\sigma_{\scriptscriptstyle 1}$ and, more important, that
$D$ no longer goes to zero when $\sigma_{\scriptscriptstyle 1}$ goes to zero. Instead, $D(\sigma_{\scriptscriptstyle 1})$ intersects
the axis $D=0$ at a shear stress $\sigma_{\scriptscriptstyle 1}$ of about 100~Pa which corresponds to $\sigma\approx 90$~Pa
(using Eq.~(\ref{e.sigmawall})).
This means that, when calculated by
$A=\sigma_{\scriptscriptstyle 1}/D$ according to Eq.~(\ref{e.D}),
$A$ varies from 407 to 185 when $\gammap$ increases from 0.4 to 15~s$^{-1}$.
This strong variation of $A$ with $\sigma$
shows that, even though the individual fits by Eq.~(\ref{vfit}) are all very good,
our data are not compatible with the flow behaviour given by Eq.~(\ref{flowcurve}) with a
single constant $A$.
In particular, Eq.~(\ref{flowcurve}) may well hold locally and for a single value of $n$, but
$A$ has to be allowed to vary with $\sigma$ in order to account for the experimental data.   
Thus, in the case
of our concentrated emulsion, the local model used in Ref.~\cite{Coussot:2002} fails to describe
the whole range of investigated shear stresses with a single exponent $n=0.4$ and a single value
of $A$.

\subsubsection{Influence of a yield stress}

In the case of the concentrated emulsion, it is very tempting
to interpret the particular value $\seuil$ of the stress where $D(\seuil)=0$ 
as a ``yield stress''. However, if we assume the following local behaviour
based on the Herschel-Bulkley equation:
\begin{equation}
\sigma(x)=\seuil+A\, \gammap(x)^n \,\,\, \hbox{\rm for} \,\,\,\sigma \ge \seuil\, ,
\label{flowcurve2}
\end{equation}
the above analysis of the velocity profiles is no longer valid. Indeed, if Eq.~(\ref{flowcurve2})
holds and if $n\ne 1$, the velocity $v(x)$ is no longer easily
integrable from Eqs.~(\ref{e.localshearrate})
and (\ref{sigmax}). Instead, for $n\ne 1$, one gets:
\begin{equation}
\frac{v(x)}{R_{\scriptscriptstyle 1}+x} = \frac{v_{\scriptscriptstyle 1}}{R_{\scriptscriptstyle 1}}+\int_{R_{\scriptscriptstyle 1}}^{R_{\scriptscriptstyle 1}+x}{\,\frac{\hbox{\rm d}r}{r}\left[
\frac{\sigma\left(\frac{R_{\scriptscriptstyle 1}}{r}\right)^2-\seuil}{A}\right]^{1/n}\,.}
\label{vtheo2}
\end{equation}

Note that we can rule out the case $n=1$ for two reasons: (i) although they do not
allow for a reliable measurement of $\seuil$, rheological data indicate
a power-law behaviour with $n\approx 0.4$ at large $\gammap$ (see Fig.~\ref{rheol75glob}),
and (ii) $n=1$ leads to velocity profiles whose curvatures varie strongly from a logarithmic
shape close to $\seuil$ to an almost linear shape when $\sigma\gg\seuil$, whereas the
curvature of our experimental profiles does not vary noticeably (see Fig.~\ref{emu75}).

We integrated Eq.~(\ref{vtheo2}) numerically and looked for a set of parameters
$v_{\scriptscriptstyle 1}$, $\seuil$, $A$, and $n$ that would describe all the different profiles.
Since $v_{\scriptscriptstyle 1}$ corresponds to slip at the rotor, the values $v_{\scriptscriptstyle 1}(\sigma)$ 
found earlier in this paper were forced into the equation.
We also set $n=0.4$ because this value seems to describe correctly the shear-thinning behaviour
of our emulsion (see Figs.~\ref{rheol75glob} and \ref{nonlin}(b)).
With only two free parameters $\seuil$ and $A$, we were then able
to fit the velocity profiles very closely so that the fits were undistinguishable
from those obtained previously with Eq.~(\ref{vfit}). However, once again, 
we could not find a single set of parameters to fit all the velocity profiles. 
$A$ and $\seuil$ had to be allowed to vary significantly with $\sigma$ so that the whole
range of investigated shear rates could be accounted for.

Due to the restricted amount of data and to the experimental uncertainty, it was
not reasonable to allow one more free parameter in the fits, namely to allow
$n$ to vary. In that case, almost any value of $n$ between 0.1 and 0.7 could be 
obtained by tuning $A$ and $\seuil$. Indeed, in the frawework of Eqs.~(\ref{flowcurve2})
and (\ref{vtheo2}), the curvature of $v(x)$ is not only controlled by $n$ but is also
very sensitive to $\seuil$.
However, if we restrict the analysis to the range $0.2<\gammap<5$~s$^{-1}$ where
$D(\sigma_{\scriptscriptstyle 1})$ appears as very linear on Fig.~\ref{prefact}(b), we showed that
$n=0.4$ and $\seuil=90\pm 10$~Pa were reasonable parameters for this range of
applied shear rates. Note that this range of shear rates corresponds to the domain
where the local flow curve of Fig.~\ref{rheol75loc} seems the most continuous.

\subsubsection{Existence of a global flow curve}

The above problems encountered during the analysis of the individual velocity
profiles raise the question of the existence of a global flow curve
for the concentrated system. Indeed,
it seems that simple approaches based on Eqs.~(\ref{flowcurve}) or (\ref{flowcurve2})
fail to describe our experimental data for the 75~\% emulsion with a
single set of parameters. The fact that the curvature of the individual profiles 
may be nicely fitted in terms of one of these two equations means that those
flow behaviours may be representative of the fluid at a local level for a given
shear rate.
However, except on a small range of $\gammap$, this kind of simple flow behaviour
does not hold globally: the parameters have to be adjusted from one shear rate to
another. This result is in agreement with measurements by Mason {\it et al.}
\cite{Mason:1996a} who report that for emulsions of droplet size 0.25~$\mu$m
and concentrated above 65~\%, $\sigma(\gammap)$ deviates from a power law at
high $\gammap$.

Thus, our results would lead to a general picture
of a sheared concentrated emulsion as a material whose flow behaviour changes
with the imposed stress due to subtle changes in its micro-structure:
\begin{equation}
\sigma(x)=\seuil(\sigma)+A(\sigma)\, \gammap(x)^{n(\sigma)}\, .
\label{flowcurve3}
\end{equation}
Indeed, even though we checked that the mean
radius of the oil droplets did not change during our experiments, the parameters
controlling the flow behaviour
such as the shear-thinning exponent $n$ may depend on the compression state of the
droplets or on their precise shape or even on the sample history.

\section{Discussion and conclusions}

In this paper, we presented local velocity measurements in a dilute and in a concentrated
emulsion using heterodyne DLS. In both systems, we found significant slippage. 
We have shown that the corrections for wall slip usually introduced in rheology are
valid for the concentrated emulsion. However, in the dilute emulsion, inertial effects
may come into play and complicate the analysis so that direct estimate of slip velocities
are required.

Once wall slip is measured, we can access interesting local rheological information.
We found that the dilute emulsion has a Newtonian behaviour and that the concentrated
system is shear-thinning with an exponent close to 0.4 that accounts well for the curvature
of the velocity profiles. It also seems that the behaviour of the concentrated emulsion
cannot be described by a single flow curve over the whole range of shear rates that we investigated.
This raises the question of the definition of a global flow curve.

We have also shown that for the smallest shear rates at which the concentrated system flows
($0.4\le\gammap\le 1$~s$^{-1}$), the flow behaviour is compatible with a yield-stress fluid
with $\seuil\approx 90$~Pa and $n\approx 0.4$. In any case, our results clearly rule out
a simple power-law behaviour with a single prefactor $A$ (see Fig.~\ref{prefact}(b)).
For $\gammap\le0.2$~s$^{-1}$, the emulsion does not
flow and behaves like a solid. This shows the existence of two different regimes: one where
the system is solid-like and one where the system flows like a yield-stress fluid.
Up to the precision of the present data, the two shear rates at which the system is solid-like
seem to correspond to shear stresses slightly above
$\seuil$ ($\sigma=100$ and 130~Pa, see Fig.~\ref{rheol75loc}). 
This raises important questions about the behaviour of a concentrated emulsion near the transition
between the two regimes.

Indeed, recent experiments \cite{Coussot:2002,Debregeas:2001}
and numerical simulations \cite{Varnik:2002pp}
on soft disordered materials have focussed on the local flow properties of yield-stress systems
like clay suspensions, industrial gels, emulsions and foams.
Such systems were shown to present inhomogeneous flow profiles when sheared close to their
yield point: the sample separates between a fluid-like phase near the rotor
and a solid-like phase near the fixed wall of a Couette cell. In some cases,
the velocity 
profiles in the fluid phase show a strong curvature very similar to that observed on our
measurements in the concentrated emulsion \cite{Coussot:2002}.
These features seem quite general since they have also been observed in colloidal suspensions
and granular pastes
\cite{Pignon:1996,Barentin:2002pp}.

Since the first models of soft glassy materials under shear \cite{Bouchaud:1995,Sollich:1997}
and triggered by recent experimental developments on aging and rejuvenation
\cite{Cloitre:2000,Ramos:2001,Viasnoff:2002},
intense theoretical effort is currently being spent on the understanding of
the yield stress phenomenon \cite{Lemaitre:2002,Fuchs:2002pp_a,Fuchs:2002pp_b} and on the
possible existence of two separate branches in the flow behaviour
\cite{Derec:2001,Berthier:2002pp}. In particular, hysteretic effects are
expected when the shear stress is controlled and the presence of shear bands is predicted
when the shear rate is imposed \cite{Picard:2002pp}.

So far, we have not been able to show that inhomogeneous flows and/or hysteretic effects
take place in the 75~\%
emulsion when sheared in our small gap geometry ($e=3$~mm for $R_{\scriptscriptstyle 1}=22$~mm). Direct visualization
of the emulsion surface as in Refs.~\cite{Mason:1996a,Princen:1985} is made impossible
by the use of a thermostated cover that prevents evaporation of our samples. Moreover, let us
recall that our velocity profiles are obtained in typically 30 minutes. The flow profiles
of Fig.~\ref{emu75_01_02} showing solid-body rotation may thus miss transient inhomogeneous
behaviours that could occur on shorter time scales such as fracture or stick-slip phenomena.

However, none of the velocity profiles presented here seems consistent with a stationary
banded flow as described in Refs.~\cite{Coussot:2002,Debregeas:2001}.
More experiments around the yield point will be needed
to conclude on this issue. In particular, in the framework of the recent theoretical approaches
of inhomogeneous flows, the existence of two points above the yield stress
where the system behaves like a solid body
should lead to inhomogeneous and possibly non-stationary flows in this region.
Future experiments will focus on this specific point. We also intend to use high-frequency ultrasound
to measure velocity profiles with an increased temporal resolution (0.01--0.1~s)
in order to capture transient and possibly non-stationary flows. 

\begin{acknowledgement}
The authors wish to thank the ``Cellule Instrumentation'' at CRPP for
building the heterodyne DLS setup and the Couette cells used in this study.
We are also very indebted to B.~Pouligny for his decisive participation to the
optical setup and for helpful physical discussions.
We thank A.~Ajdari, C.~Barentin, L.~Bocquet, C.~Gay, F.~Molino,
P.~Olmsted, G.~Picard, D.~Roux, and C.~Ybert for fruitful
discussions on the interpretation of our data. 
\end{acknowledgement}

\end{document}